\documentclass[prc,amsfonts, amssymb, amsmath, nofootinbib,twocolumn,superscriptaddress]{revtex4-1}
\usepackage[utf8]{inputenc}
\usepackage{graphicx}
\usepackage{mathrsfs}
\usepackage{makecell}
\usepackage[caption=false]{subfig}
\usepackage{color}
\usepackage[dvipsnames]{xcolor}
\usepackage{comment}
\usepackage{ulem}
\newcommand{\mtc}[1]{\mathcal{#1}}

\newcommand{\clg}{\color{black}}

\usepackage{textcase}   

\begin{document}

\title{ \boldmath{$np \leftrightarrow d\gamma$} reactions calculated up to $E_{\gamma}=20$ MeV}
\date{\today} 

\begin{abstract}
We calculate the electromagnetic dipole transition cross sections for the $np \rightarrow d\gamma$ and $ d\gamma \rightarrow np$ reactions over a broad range of energies.~We use the LENPIC nucleon-nucleon interaction obtained from chiral effective field theory ($\chi$EFT) up to next-to-next-to-next-to-next-to-leading order (N4LO) and effective electromagnetic dipole transition operators obtained from the same $\chi$EFT up to N2LO.  Our results agree with existing experiments.~We get results at energies for which experimental data and/or modern theoretical calculations have not been reported. \color{black} In this study, we utilize a new approach, namely, our adaptation of the Efros [V.~D.~Efros, Phys. Rev. C \textbf{99}, 034620 (2019)] method that is prospective for future many-body applications in calculations of bound and continuum state wave functions.  
\end{abstract}
\author{Mamoon A. Sharaf}
\affiliation{Institute of Modern Physics, Chinese Academy of Sciences, Lanzhou 730000, China}
\author{Weijie Du}
 \email{Corresponding author: duweigy@gmail.com}
\affiliation{Institute of Modern Physics, Chinese Academy of Sciences, Lanzhou 730000, China}
\author{Andrey M. Shirokov}
\affiliation{Skobeltsyn Institute of Nuclear Physics, Lomonosov Moscow State University, Moscow 119991, Russia}
\maketitle

\section{Introduction}
\label{secI}

\hspace{1.5mm} The reactions~$np \rightarrow d\gamma$ and $ d\gamma \rightarrow np$ are key processes in primordial nucleosynthesis and play a key role in the study of nuclear astrophysics \cite{ArenhovelSanzone1991, Ian, Pisanti, Cooke}, particularly at low energies.  Experimental data for these reactions are sparse at low energies \cite{BarnesExp, CoxExp, GhemoExp, SkopikExp, AhrensExp,CokinosExp, BirenbaumExp, StiehlerExp, BernabeiExp, MorehExp, MichelExp, GraeveExp, SuzukiExp, NagaiExp, TomyoExp, HaraExp, AndersonExp, ChenExp}.~Therefore, theoretical studies with well-established predictive power are needed.~Prior theoretical studies have been applied to the reactions that we investigate here and have invoked a variety of methods including phenomenological approaches~\cite{Partovi1964, Cambi1982} utilized in Refs.~\cite{BirenbaumExp, Partovi1964, BernabeiExp}, the use of the Paris~\cite{Paris1980} and Bonn~\cite{Bonn1986} NN potentials in Ref.~\cite{FinkExp}, the chiral Lagrangian approach for obtaining exchange vector currents in nuclei within chiral perturbation theory~\cite{TSPark1996}, the pionless effective field theory (EFT) \cite{ChenSavage,Rupak, Ando2006, Richardson2025}, a hybrid approach that utilizes both chiral EFT ($\chi$EFT) currents and phenomenological potentials~\cite{YHSong2009}, and Lattice QCD~\cite{ Beane2015}.  It is important to study those reactions using more modern $\chi$EFT interactions~\cite{ Epelbaum2015, Reinert2018} and operators~\cite{Piarulli2013}.  Such theoretical investigations were carried out in Refs.~\cite{Acharya2022} and \cite{DuWeijieHORSE}.  

In this work, we further explore these reactions using modern interactions and operators derived from $\chi$EFT.  We present results for the reaction $np \rightarrow d\gamma$ at higher energies using the same $\chi$EFT NN interaction~\cite{Epelbaum2015}  up to N4LO and electromagnetic operators~\cite{SPal2023} up to N2LO as in our previous work~\cite{DuWeijieHORSE}. We also provide results for the inverse reaction $d\gamma \rightarrow np$. To our knowledge, some of the results reported here are the first theoretical descriptions of experimental cross sections.  We further note that, in several cases, either experimental measurements or theoretical results have not been reported previously.

It is also important to note that we verify and demonstrate the applicability of a new method for describing the nuclear continuum in this work.~This method is promising for extending the \textit{ab initio} (from first principles) no-core shell model (NCSM) \cite{NCSMPRL, NCSMProg} to the continuum spectrum and for developing an \textit{ab initio} many-body reaction theory.

We note that various \textit{ab initio} methods have been extensively developed and applied with the aim of describing the structure of atomic nuclei.~These applications are a modern trend in the theory of light nuclei that are well-developed and have been successfully used to calculate various bound systems.~A number of \textit{ab initio} methods have been developed and successfully applied to elastic scattering problems with $A > 4$ nucleons.~Those include the no-core Gamow shell model (NCGSM) \cite{Papadimitriou}, the resonating group method in combination with the NCSM (RGM/NCSM) \cite{QuagOne}, the NCSM with continuum (NCSMC)~\cite{BaroniOne}, and the single-state harmonic oscillator representation of scattering equations (SS-HORSE)~\cite{AMShirokovSSHORSE2016}, which is a simplification of the HORSE method~\cite{Zaitsev1998,BangHORSE}.~We employed the HORSE method in our previous work on $np \rightarrow d\gamma$~\cite{DuWeijieHORSE}.~The NCGSM, the RGM/NCSM, the NCSMC, and the SS-HORSE method are all based on the NCSM, see Ref.~\cite{CWJohnson} for an overview of the different \textit{ab initio} techniques.

Despite the successes of these approaches, there are major difficulties in applying them to describe \textit{ab initio} scattering states and reactions in systems with \mbox{$A > 4$} \cite{CWJohnson}.~In particular, the NCGSM, the RGM/NCSM, and the NCSMC are computationally expensive while the HORSE method is impractical for \textit{ab initio} many-body applications due to the prohibitive computational cost.~On the other hand, the SS-HORSE method is simple and can be used to obtain some important observables in light nuclear systems.  However, it is unable to calculate the continuum wave function.~Without the continuum wave function, other important physical observables such as electromagnetic transition cross sections are not obtainable.~Moreover, it is unclear how to extend this approach to multi-channel problems.

To address these issues, the Hulthén–Kohn method was modified by Efros \cite{VDEfros} in a way that makes it convenient for extending the NCSM to the continuum spectrum.~In our previous work~\cite{MethodsPaper2024}, we adapted this approach by incorporating the ideas of the HORSE method and employing the oscillator basis in all equations.  This facilitates \textit{ab initio} many-body calculations of scattering wave functions and $S$-matrix poles using a limited set of short-range functions (SRFs) and interaction truncations accessible in modern NCSM calculations of light nuclei.~We will refer to our adaptation of the Efros method simply as the ``Efros method".

As the first application, we employ the Efros method to calculate the reactions~$np \rightarrow d\gamma$ and $ d\gamma \rightarrow np$ using a modern $\chi$EFT interaction and corresponding $\chi$EFT electromagnetic operators.~We extend our results \cite{DuWeijieHORSE} to higher energies, including some energies where, to our knowledge, experimental data or modern theoretical calculations are unavailable, and, in some cases, both are lacking.~We use this problem where the exact wave functions can be obtained by other methods, e.\,g., by a direct integration of the Schrödinger equation or by the HORSE method where a complete convergence can be achieved, as a further test of the Efros method, which we plan to utilize in future many-body applications.  

For the $np \rightarrow d\gamma$ reaction, we consider the c.m. energies $E$ ranging from astrophysical energies to $17.78$ MeV that corresponds to $E_{\gamma}=20$ MeV.~For these energy ranges, only the magnetic ($M1$) and electric ($E1$) dipole transitions contribute significantly.~As in Ref.~\cite{DuWeijieHORSE}, we use the LENPIC N4LO interaction regularized by the semi-local coordinate space regulator $R=1$ fm~\cite{Epelbaum2015} and the $M1$ transition operator as derived from the same $\chi$EFT up to N2LO in Ref.~\cite{SPal2023}.  We use only the naive $E1$ operator since the NLO and N2LO corrections for $E1$ were shown to be negligible~\cite{SSarkerThesis}.

\section{Basic Equations}

\hspace{1.5mm} The normalized scattering wave function for the $np$ system at \mbox{c.m. energy} $E$ with reduced mass \mbox{$m=469.46 \hspace{1mm} \textrm{MeV}/c^{2}$} is parameterized in the uncoupled partial waves as
  \begin{multline}
\label{eq2Smat}
\langle \vec r|\Psi^{\Gamma} \rangle=
\frac{i}{2}[\eta^-(\vec{r})
-\eta^+(\vec{r})\, \mathbb{S}(E)]
\\
+\sum_{q=1}^{v}b_{q}(E)\,\beta_{q}(\vec r)
\end{multline}
and in the coupled partial waves with ingoing spherical wave in the partial wave $i$ as
\begin{multline}
\label{eq3Smat}
\langle \vec r|\Psi^{\Gamma_{i}} \rangle=\frac{i}{2}[\eta^-_i(\vec{r})
-\sum_{j=1}^{2}\eta^+_j(\vec{r})\, \mathbb{S}_{ji}(E)]
\\
+\sum_{q=1}^{v}b_{q}^{i}(E)\,\beta_{q}(\vec r).
\end{multline}   
Here, $\vec r$ is the distance between the proton and the neutron and $i=1,2$.~We note that equations involving coupled partial waves are formally equivalent to those of coupled channels.  The multi-index $\Gamma=\{J, S, l\}$ defines a particular partial wave with quantum numbers $J$ which is the total angular momentum, $l$, the orbital angular momentum of the relative motion, and $S$, the total spin.  $\eta^\pm_i(\vec{r})$ describe the long-range behavior of the scattering wave functions, it is convenient to express them through infinite series of the oscillator functions as discussed in Ref.~\cite{MethodsPaper2024} and in Appendix~\ref{AppendixA}, \color{black}  and $\beta_{q}(\vec r),~q=1,...,v$, are $v$ linearly-independent SRFs vanishing as \mbox{$r \rightarrow \infty$} which are introduced to describe the wave functions at short distances.

For the uncoupled partial waves, the set of $v$ energy-dependent coefficients~$b_{q}(E),~q=1,...\:,v$, together with the $S$-matrix $\mathbb{S}(E)=e^{2i\delta_{l}(E)}$, where $\delta_{l}(E)$ is the scattering phase shift, comprise a set of $v+1$ unknowns.~In the case of coupled partial waves, the set of $v$ energy-dependent coefficients~$b_{q}^{i}(E),~q=1,...\:,v$, together with two $S$-matrix elements~$\mathbb{S}_{1i}(E)$ and $\mathbb{S}_{2i}(E)$, comprise a set of~$v+2$~unknowns in each incoming partial wave, \mbox{$i=1,2$}.  Therefore, there is a total of $2v+4$ unknowns.~However, because of the unitary structure of the $S$ matrix, there are only $2v+3$ independent unknowns.~Those unknowns are obtained by our adaptation of the Efros method as discussed in Ref.~\cite{MethodsPaper2024}.~The only difference with Ref.~\cite{MethodsPaper2024} is that we are using the $K$-matrix representation to obtain solutions that guarantee $S$-matrix unitarity for coupled partial waves.  We denote the $K$ matrix by $\mathbb{K}$, where the $K$ and $S$ matrices are related as
\begin{equation}
\label{eq11}
\mathbb{S}=\frac{1+i\mathbb{K}}{1-i\mathbb{K}}.
\end{equation} 
For coupled partial waves, we impose symmetry by setting $\mathbb{K}_{12}(E)=\mathbb{K}_{21}(E)$.

The deuteron bound state energy $E_{B}$ and wave function $\langle \vec r|\Psi^{J=1, S=1} \rangle$ in the $^3SD_{1}$ partial wave can be found by locating the $S$-matrix pole, see Ref.~\cite{MethodsPaper2024} for details.  We employ $S$-matrix pole location because it provides the deuteron wave function as an infinite expansion in the oscillator basis providing the correct asymptotic tail.  This is very important for calculating capture cross sections at low energies.  In general, this is important for weakly-bound systems, e.g., for calculations of energies and radii in nuclear halo systems.  Moreover, we get a much faster convergence of energies and other observables if we use the method of $S$-matrix pole location as compared to the direct diagonalization of the Hamiltonian matrix.

 We use the oscillator basis to construct the Hamiltonian of the $np$ system.  The oscillator basis representation for the Hamiltonian is natural and convenient in many-body \textit{ab initio} nuclear physics applications such as in the NCSM.  We denote the oscillator function by
\begin{equation}
\label{HOFn1}
\phi_{n j}(\vec r)  \equiv \phi_{nl_{j}}\!(\vec r) = R_{nl_{j}}\!(r)\,\mathscr{Y}_{J_{j} M_{J_{j}}}^{l_{j} S_{j}}(\Omega_{\hat{r}}),
\end{equation}
where $R_{nl_{j}}\!(r)$ is the radial component, $n$ is the radial quantum number, and $l_{j}$ is the orbital angular momentum of relative motion in the channel $j$.   The radial component $R_{nl}\!(r)$ is parameterized by the oscillator length parameter $b=\sqrt{\hbar /\ \hspace{-1mm} m\Omega}$, where $\Omega$~is the oscillator frequency.~We take $\hbar \Omega=28$ MeV throughout this work.~It was tested that the calculations in the range \mbox{$\hbar \Omega=20-30$ MeV} provide reasonable convergence and the respective results are consistent. \color{black}  $ \mathscr{Y}_{J M_{J}}^{lS}(\Omega_{\hat{r}})$ is the generalized spherical harmonic~\cite{Varsha}, and $M_{J}$ is the total angular momentum projection.

We use the Hamiltonian 
\begin{multline}
\label{eq21Hamiltonian}
H=T+\tilde{V}={\clg\sum_{i }}\:\sum_{n,n'=0}^{\infty}|\phi_{n i}\rangle T^{l_{i}}_{nn'}\langle\phi_{n'i}| 
\\
+{\clg\sum_{i,j}}\sum_{n=0}^{\mtc{N}_{i}}\sum_{n'=0}^{\mtc{N}_{j}}|\phi_{ni}\rangle \sigma_{\mtc{N}_{i}}^{n} V_{n i, n'j} \sigma_{\mtc{N}_{j}}^{n'} \langle\phi_{n'j}|.
\end{multline}
Here, $T^{l_{i}}_{nn'}$ are the elements of the tridiagonal kinetic energy matrix and~$V_{n i, n'j}$ are the interaction matrix elements.~As in the HORSE method~\cite{Zaitsev1998,BangHORSE}, we use the infinite kinetic energy matrix in order to allow for scattering and approximate the potential energy $V$ by a finite matrix in the oscillator basis.~This is justified because the potential energy matrix elements decrease with~$n$ and/or~$n'$ and can be neglected at large~$n$ and/or~$n'$ when compared with the non-zero kinetic energy matrix elements that are increasing with radial quantum numbers $n$.~For the $np$ system, we define the truncation boundaries of the potential energy matrix by the maximal oscillator quanta $N_{\max}$.  $\mtc{N}_{i}$ is the truncation boundary of the potential energy matrix in scattering channel $i$.~In particular, ${2n+l_{i} }\leq N_{\max}$, $n=0,1,...\,,\mtc{N}_{i}$.  We multiply the interaction matrix elements $V_{n i, n'j}$ by $\sigma_{\mtc{N}_{i}}^{n}$ and $\sigma_{\mtc{N}_{j}}^{n'}$ given by
\begin{equation}
\label{eq20}
\sigma_{\mtc{N}}^{n}=
\frac{1-e^{-\left(\!a\frac{n-\mtc{N}-1}{\mtc{N}+1}\!\right)^{\hspace{-.7pt}2}}}{1-e^{-a^{2}}}
\end{equation}
 in order to improve convergence by smoothing the potential truncation as suggested in Refs.~\cite{HungOne, HungTwo}.  Here, $a$ is a dimensionless parameter.  We take $a=7.5$ throughout this work (we have checked different values of $a$ and found out that $a=2.5-10$ is an optimal range from the point of view of convergence).

For the uncoupled partial waves, we use the SRFs 
\begin{gather}
\label{eq22}
 \beta_{q}(\vec r)= 
\ \begin{cases} 
 \beta_{\frac{q+1}{2}}^{\textrm{Efn}}(\vec r), & q=1,3,...,v-1+\textrm{mod}(v,2),  \\ 
     \phi_{\mtc{N}-(\frac{q}{2})}(\vec r), & q=2,4,...\, ,v-\textrm{mod}(v,2).
   \end{cases}
\end{gather}
We note that these SRF choices suggest faster convergence than those discussed in Ref.~\cite{MethodsPaper2024} for the uncoupled partial waves.  For the coupled partial waves, we use the SRFs
\begin{equation}
\label{eq25}
{\beta_q(\vec r)} 
=\beta_{q}^{\textrm{Efn}}(\vec r), \quad q=1,...,v.
\end{equation}

Here, $\beta_{q}^{\textrm{Efn}}(\vec r)$ are the eigenfunctions of the truncated Hamiltonian
\begin{equation}
\label{eq26}
H^{tr}=
{\sum_{i,j}} \sum_{n=0}^{\mtc{N}_{i}}\sum_{n'=0}^{\mtc{N}_{j}}
|\phi_{ni}\rangle \langle \phi_{ni} |H|\phi_{n'j}\rangle \langle\phi_{n'j}|.
\end{equation}
We denote the lowest-lying state by $\beta_{1}^{\textrm{Efn}}(\vec r)$, the first excited state by $\beta_{2}^{\textrm{Efn}}(\vec r)$, etc., in a given partial wave.  In some cases, it is favorable to use a different SRF selection due to jumps in phase shifts that can occur in coupled partial waves at certain energies within the Efros method~\cite{MethodsPaper2024}.~In particular, for the $^3PF_{2}$ partial wave, we use for $v=4$, $N_{\textrm{max}}=20$, $\hbar \Omega=28$ MeV, and $a=7.5$ the selection $q=1,2,6,4$ (i.e., for a given partial wave, we use the lowest-lying state, the first excited state, the fifth excited state, and the third excited state) to avoid a jump that occurs at around $E=0.5$ MeV in the $^3PF_{2}$ partial wave.

 We expand the eigenfunction SRFs in a finite series of oscillator functions $\phi_{q'i}(\vec r)$,
\begin{equation}
\label{eq27}
\beta_{q}^{\textrm{Efn}}(\vec r)=\sum_{i}\sum_{q'=0}^{\mtc{N}_{i}}a_{qq'}^{l_{i}}\,{\phi_{q'i}(\vec r)}, \quad q=1,...,v,
\end{equation}
where the coefficients $a_{q q'}^{l_{i}}$ are obtained from the diagonalization of the matrix $H^{tr}$.~We note that the Hamiltonian structure of $H^{tr}$ in which both the kinetic energy and interaction matrix elements are truncated, is conventionally employed in NCSM calculations.  For the uncoupled partial waves, we exclude the summation over $i$ in Eq.~\eqref{eq27}.

  As was mentioned in Ref.~\cite{MethodsPaper2024}, the Efros method
is equivalent to the HORSE approach if the set of SRFs includes the ``complete" set of eigenfunctions (i.e., all the eigenfunctions  of $H^{tr}$).~However, because we have in mind future applications in many-body systems in combination with the NCSM in which only the lowest-lying eigenstates and limited $N_{\textrm{max}}$ are accessible, we prefer to use only the lowest-lying eigenfunctions in a given partial wave in calculations.  Therefore, we try to use both a limited interaction truncation $N_{\textrm{max}}$ and a small number of low-lying SRFs $v$.~\color{black}  We have checked various combinations of SRFs and have found that the sets of SRFs that we utilize provide excellent convergence with increasing $v$. 

We write the total capture cross section of the reaction $np \rightarrow d\gamma$ as
\begin{multline}
\label{eq1}
\sigma^{\textrm{cap}}=\frac{1}{4}\frac{16 \pi}{9u}\frac{k_{\gamma}^{3}}{k^{2}}\bigg[\sum_{i}|\langle \Psi^{J=1,S=1}||\mathcal{M}^{M}_{1}||\Psi^{\Gamma_{i}}\color{black}\rangle |^{2}
\\
+\sum_{i}|\langle \Psi^{J=1,S=1}||\mathcal{M}^{E}_{1}||\Psi^{\Gamma_{i}}\color{black}\rangle  |^{2}\bigg] 
\end{multline}
 following Ref.~\cite{Descouvemont2005}.~Here, the factor of $\frac{1}{4}$ accounts for the fact that, for unpolarized beam and target, we average over the initial states. $u=\hbar k/ m c$ is the relative velocity of the $np$ system.~\mbox{$k=\sqrt{2mE} / \hbar$} is the relative momentum of the $np$ system and $k_{\gamma}=E_{\gamma}/ \hbar c=~(E+|E_{\textrm{B}}|)/\hbar c$ is the photon momentum.~$\langle \Psi^{J=1,S=1}||\mathcal{M}^{M}_{1}||\Psi^{\Gamma_{i}}\rangle$ and $\langle \Psi^{J=1,S=1}||\mathcal{M}^{E}_{1}||\Psi^{\Gamma_{i}}\rangle $ are the reduced $M1$ and $E1$ transition matrix elements, respectively, where each $\Psi^{\Gamma_{i}}$ is the initial $np$ scattering states defined by the ingoing spherical wave in the partial wave $i$. 
  
We calculate the total photodisintegration cross section from the total capture cross section via detailed balance
\begin{equation}
\label{eq34}
\sigma^{\textrm{photodis}}=\frac{2}{3}\frac{k^{2}}{k_{\gamma}^{2}}\sigma^{\textrm{cap}}
\end{equation}
as was done, e.\,g., in Ref.~\cite{ChenSavage}, where the factor of $\frac{2}{3}$ is a statistical factor to account for the fact that there are three deuteron polarizations and two photon polarizations in the initial state in the $d\gamma \rightarrow np$ reaction.

\section{Results}
\label{sec2}

\hspace{1.5mm } We calculate the bound state by searching for the corresponding $S$-matrix pole in the coupled $^3{SD_{1}}$ partial waves.~We find that~$v=118$~for an interaction truncation $N_{\textrm{max}}=120$ [see Eq.~\eqref{eq26}]~gives sufficiently accurate results.~We therefore use this truncation in all calculations.~We obtain \mbox{$E_{B}=-2.2232$ MeV}.  Relativistic corrections are needed~\cite{EpelbaumCorrespondence} in order to get better agreement with the experimental value of \mbox{$E_{B}=-2.224575(9)$~MeV~\cite{VanDerLeun1982}.} We note that we require a large interaction truncation $N_{\textrm{max}}$ because of the weak binding, a large $\hbar \Omega$ value, and the need for accurate wave function tails for the calculations of electromagnetic cross sections at astrophysical energies, where the asymptotic behavior of the wave function is important.

  We find that our calculated quadrupole moment \mbox{$Q=0.2723 \hspace{1mm} \textrm{fm}^{2}$} compares well with the value supported by the employed $\chi$EFT interaction~\cite{EpelbaumCorrespondence}.~To get a better agreement with the experimental value of~\mbox{$Q=0.2859(3) \hspace{1mm} \textrm{fm}^{2}$} \cite{Bishop1979}, one needs to account for relativistic corrections and meson-exchange contributions, as is stated in Ref.~\cite{Epelbaum2015}.

  We also find that our calculated $s$-wave asymptotic normalization coefficient \mbox{$A_{s}=0.8846 \hspace{1mm} \textrm{fm}^{-\frac{1}{2}}$} agrees with the chiral N4LO result~\cite{EpelbaumCorrespondence} that is consistent with the experimental value of \mbox{$A_{s}=0.8846(9) \hspace{1mm} \textrm{fm}^{-\frac{1}{2}}$~\cite{Ericson1983}}.~Due to the inaccuracy in the $d$-wave asymptotics, the calculated asymptotic $D/S$ state ratio \mbox{$\eta=0.0176$} deviates from the chiral N4LO result~\cite{EpelbaumCorrespondence} that is consistent with the experimental value of \mbox{$\eta=0.0256(4)$ }\cite{RodningKnutson1990}.~To improve this result, one needs larger truncations and/or a smaller $\hbar \Omega$ value.~However, the internal part of the wave function is accurately described.~We also note that the $d$-wave asymptotics plays a negligible role in the observables of interest and do not contribute to more than 0.2\% to the total electromagnetic transition cross section at the energy range of interest (see Appendix~\ref{AppendixB} for the behavior of the bound state wave function amplitudes).

             \begin{figure}[t!]
\subfloat[]{%
  \includegraphics[width=1\columnwidth]{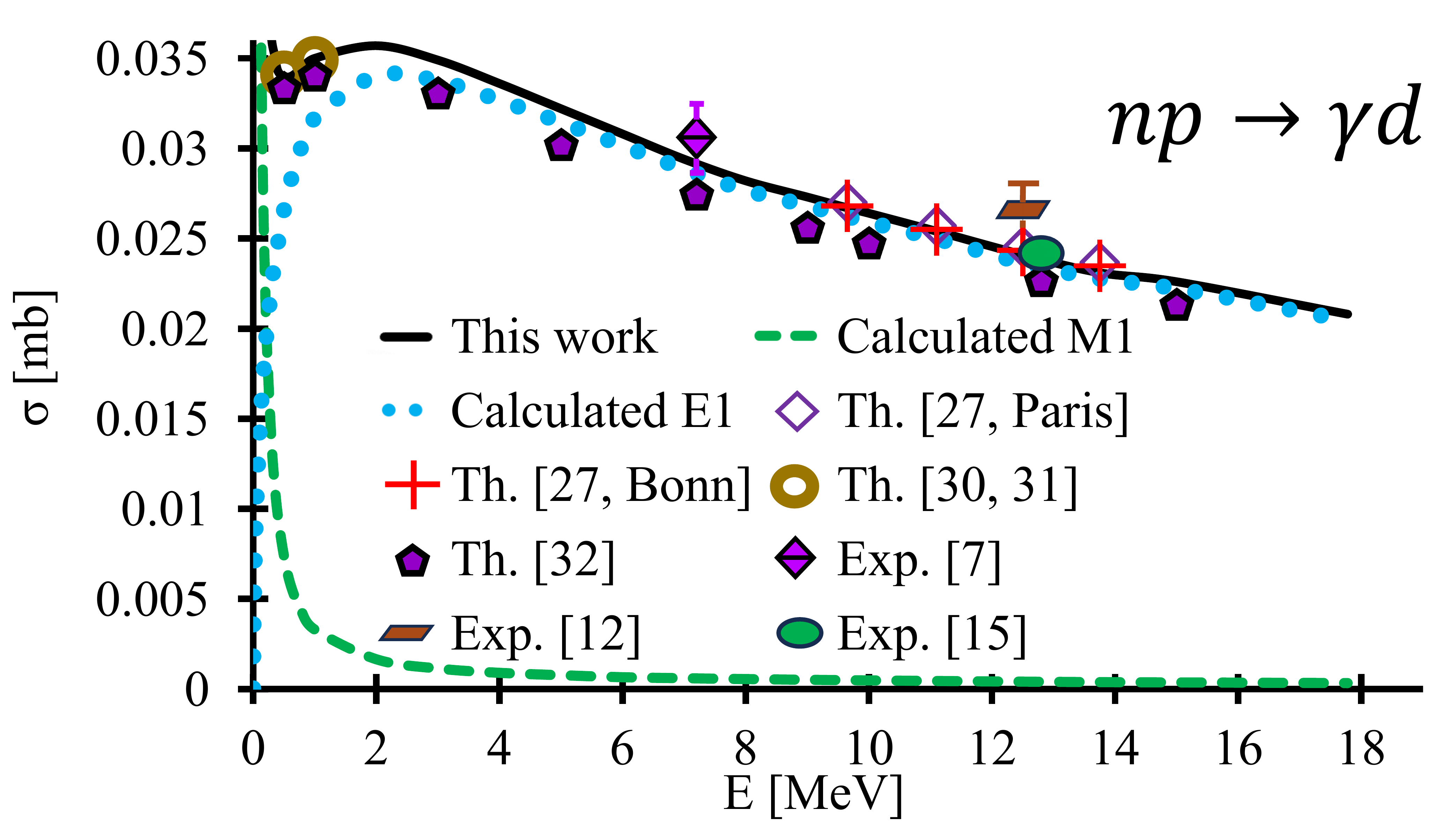}%
  \label{Fig1CAPCS}
}\hfill
\subfloat[]{%
  \includegraphics[width=1\columnwidth]{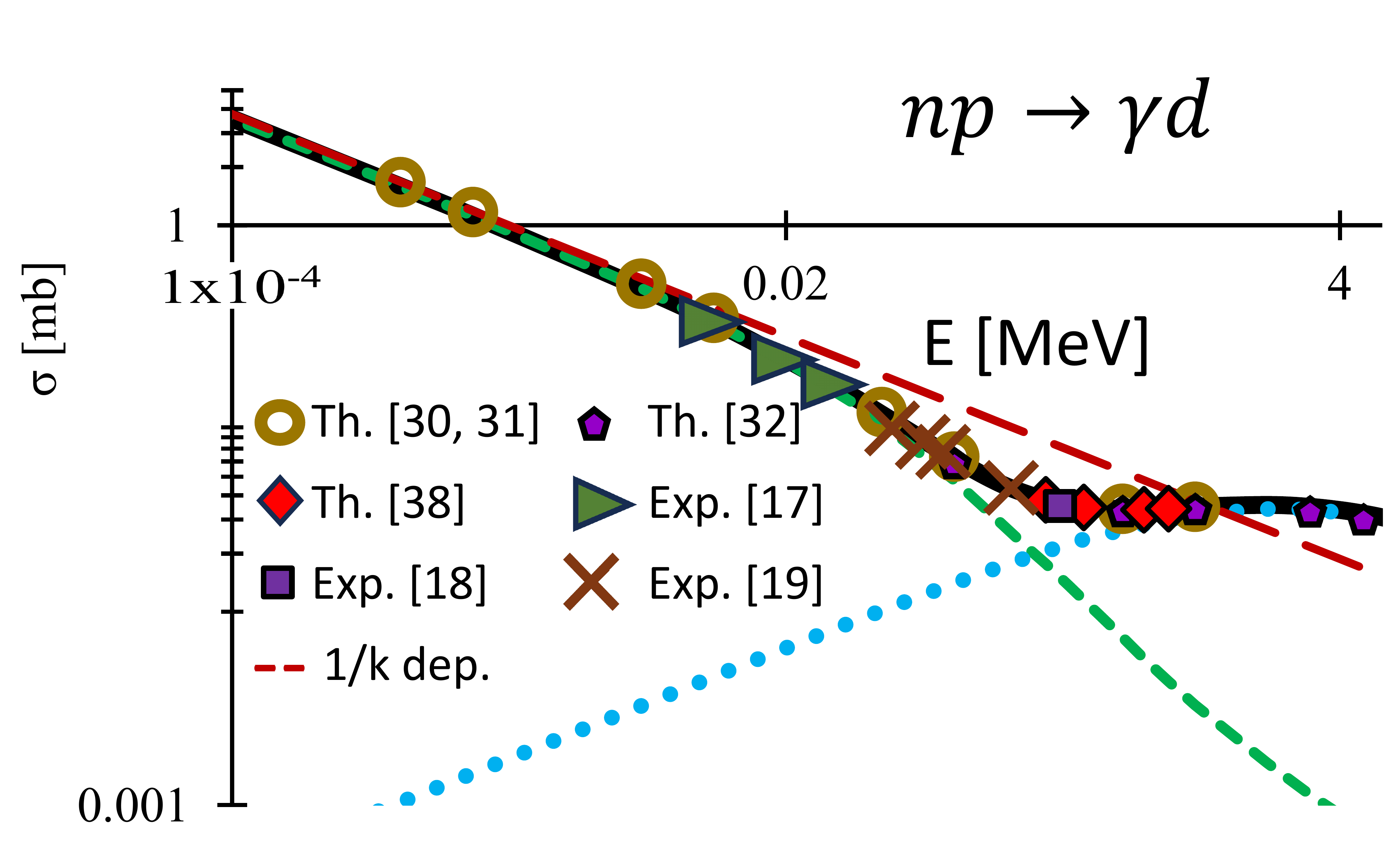}%
  \label{Fig1InsetCAPCSInset}
  }\hfill
\subfloat[]{%
  \includegraphics[width=1\columnwidth]{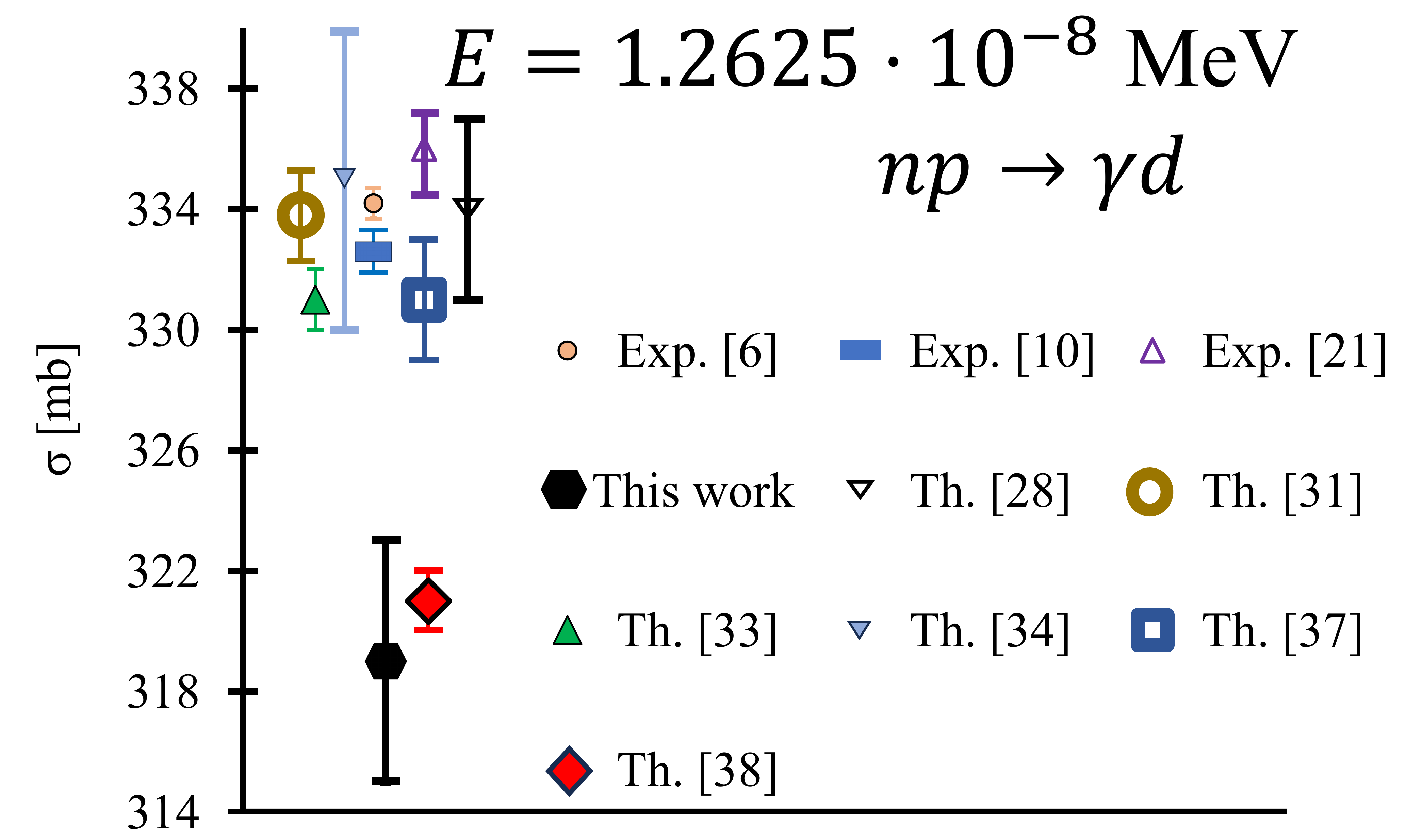}%
    \label{Fig2PHOTODISCS}
}\hfill
\caption[]{  a), b) Capture cross section plotted as a function of $E$.~The  separate $M1$ and $E1$ contributions are plotted and the results are compared with experimental data and with those from other theoretical calculations with different $NN$ interactions as well as with the $1/k$ dependence of the $M1$ capture cross section at low energies.  Most of the calculated uncertainties
are not visible at this scale. c) Capture cross section at \mbox{$E=1.2625 \cdot 10^{-8}$ MeV} compared with different experimental data and theoretical calculations.    }
 \label{figCS1}
\end{figure}

             \begin{figure}[t!]
  \includegraphics[width=1\columnwidth]{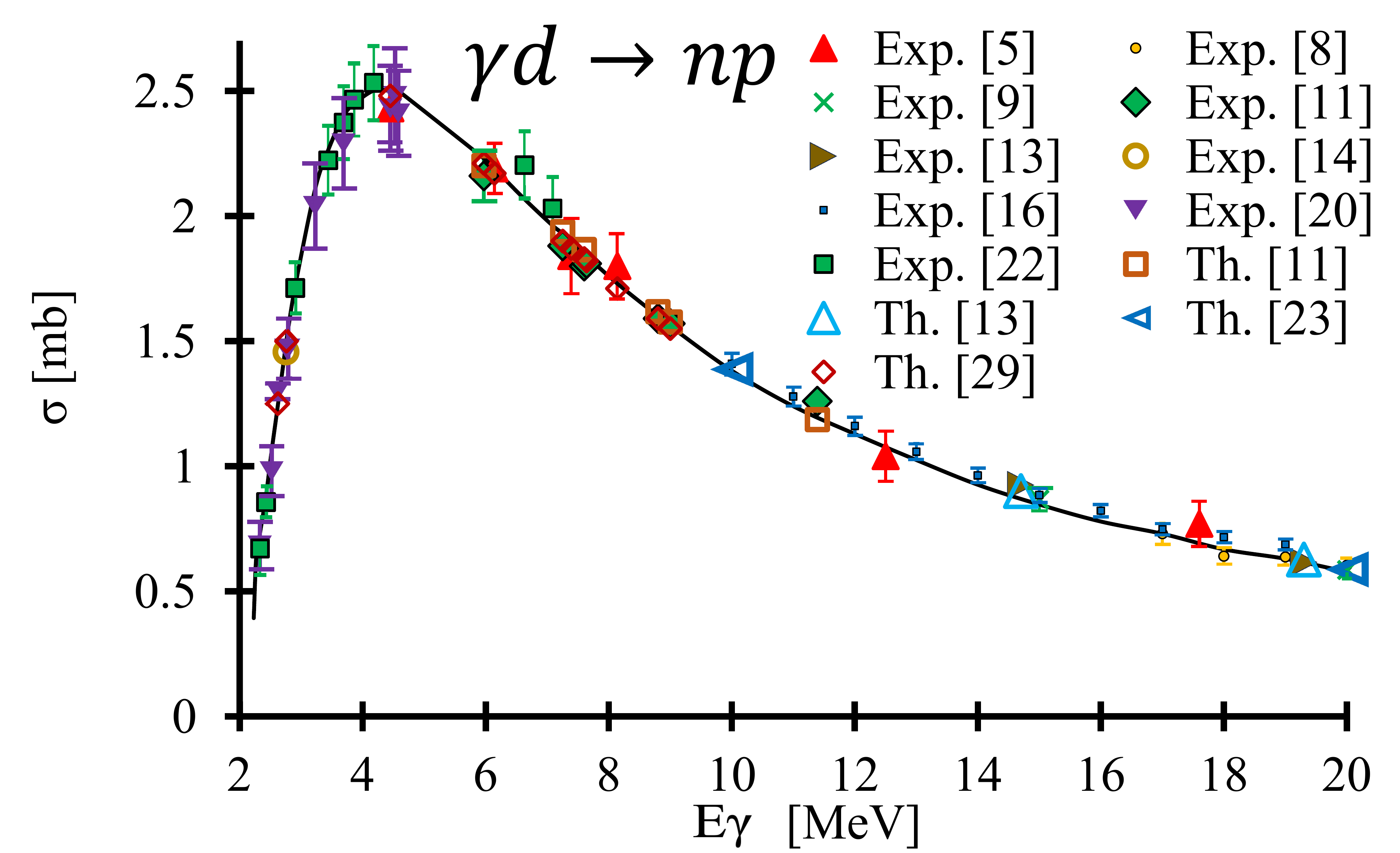}%
\caption[]{   Photodisintegration cross section plotted as a function of $E_{\gamma}$.  For comparison, the experimental data as well as other theoretical results are also shown.   }
 \label{figCS2}
\end{figure}

We present the electromagnetic transition cross section results in Figs.~\ref{figCS1} and~\ref{figCS2}.  Numerical results are presented in Tables~\ref{tab5} and \ref{tab6} in Appendix~\ref{AppendixC}.~Note that we include cross sections at energies where, to our knowledge, no theoretical or experimental results are available.~For example, we find sparse experimental data for the capture cross section over the range of $E$ from $0.3$ to $17.78$ MeV, and no theoretical results for the photodisintegration cross section over the range of $E_{\gamma}$ from  $15$ to $19$ MeV.

We estimate the uncertainty of the electromagnetic transition cross section due to the truncation of $v$ and $N_{\textrm{max}}$ for a given partial wave by checking the convergence of the cross section with $v$ and $N_{\textrm{max}}$.~We find the smallest truncations where the results converge.~We arrive at a total uncertainty in the calculated cross section by adding the uncertainties in the individual partial-wave contributions.~We find that within the calculated uncertainties, our results agree with the exact ones.  

For a given energy, we obtain results corresponding to the $^1S_{0}$, $^3P_{0}$, $^3P_{1}$, and $^3PF_{2}$ partial-wave contributions to the total capture cross section.~For the $M1$ contribution to the total electromagnetic transition cross section, we consider only the $^1S_{0}$ partial wave.~This is because the $^1S_{0}$ partial wave dominates the $M1$ contribution at $E \leq 8$ MeV.~Beyond $E=8$ MeV, the $M1$ contribution becomes negligible compared to the $E1$ contribution.~For the $E1$ transition at the energies below $E=17.78$ MeV, only the $^3P_{0}$, $^3P_{1}$, and $^3PF_{2}$ partial waves provide non-negligible contributions.~From the total capture cross section, we obtain the total photodisintegration cross section via Eq.~\eqref{eq34}.

We find that for scattering states, using $N_{\textrm{max}}=50$ and $N_{\textrm{max}}=20$ for the $M1$ and $E1$ contributions, respectively, provides reasonable truncations that assure the accuracy of the cross sections to within 1\% in most cases and to within 2\% in a few cases only.~For the $M1$ calculation, $N_{\textrm{max}}=50$ is significantly smaller than the $N_{\textrm{max}}=180$ used in Ref.~\cite{DuWeijieHORSE}, but is still well above the number of quanta accessible for modern many-body NCSM calculations whereas for the $E1$ case, $N_{\textrm{max}}=20$ is more accessible to the many-body NCSM calculations of light nuclei.~The large $N_{\textrm{max}}$ values needed to calculate the $M1$ transitions are due to the weakly-bound deuteron, whose asymptotics play a significant role in calculations at low energies.

For the $M1$ capture cross section, we find that using $v$ ranging from $9$ to $15$ SRFs gives converged results.~For the $E1$ capture cross section, $v$ ranging from $2$ to $6$ gives convergence with the exception of the $^3PF_{2}$ contribution at $E=17.78$ MeV, where $v=10$ SRFs are needed.  $v$ values are listed in Tables~\ref{tab5} and \ref{tab6} in Appendix~\ref{AppendixC}.

We compare our results with the experimental data~\cite{CoxExp, GhemoExp, CokinosExp, StiehlerExp, MichelExp, SuzukiExp, NagaiExp, TomyoExp, AndersonExp}.~We agree with experiment everywhere except at \mbox{$E=1.2625 \cdot 10^{-8}$~MeV}, where we have a discrepancy at the order of a few percent with the results of Refs.~\cite{CoxExp, CokinosExp, AndersonExp}.  At $E=0.05512$ MeV and $12.5$ MeV, we also have similar discrepancies with the results of Refs.~\cite{TomyoExp} and \cite{StiehlerExp}, respectively.  The reason for the discrepancy at $E=1.2625 \cdot 10^{-8}$~MeV is that achieving a percent-level uncertainty with experiment requires a current operator at a higher chiral order than what we consider here.  In particular, it was noted in Ref.~\cite{Piarulli2013} that a chiral order of up to N3LO is needed for a percent-level test of agreement with experiment.

Our calculations generally agree with those of other theoretical calculations with a few exceptions.~In particular, we have discrepancies at the order of a few percent with the results of Refs.~\cite{Rupak} and \cite{Ando2006}, which use pionless EFT, at energies~up~to~\mbox{$E=0.05$ MeV} (Ref.~\cite{Rupak} uses the experimental value~\cite{CoxExp} of the capture cross section at \mbox{$E=1.2625 \cdot 10^{-8}$}~MeV as input).~These discrepancies diminish to the order of one percent at \mbox{$E=0.1$~MeV}.  We also have discrepancies at the level of a few percent with the results of Ref.~\cite{Richardson2025} above $E=0.5$ MeV.  We note however, that Ref.~\cite{Richardson2025} calculates the capture cross section up to NLO within the velocity-renormalized pionless EFT; those discrepancies are expected to be reduced if N2LO corrections are included~\cite{Richardson2025}.

In addition, we have discrepancies at the level of a few percent with the capture cross section results at \mbox{$E=1.2625 \cdot 10^{-8}$ MeV} of Refs.~\cite{TSPark1996, YHSong2009, Beane2015, Piarulli2013}.~Reference~\cite{TSPark1996} uses the chiral Lagrangian approach to derive electromagnetic currents up to N3LO within chiral perturbation theory using the heavy-baryon formalism.~Reference~\cite{YHSong2009} utilizes the same formalism, but accounts for some contributions due to contact terms ignored in Ref.~\cite{TSPark1996} and, as a result, the cutoff dependence of the $np$ capture cross section is removed.~Reference~\cite{Beane2015} uses Lattice QCD. Reference~\cite{Piarulli2013} uses $\chi$EFT to derive currents up to N3LO within time-ordered perturbation theory.  The results of Refs.~\cite{TSPark1996, Ando2006,YHSong2009, Beane2015, Piarulli2013} have a percentage-level agreement with experiment.~Our discrepancies with the results of these theoretical calculations are expected to diminish if we include a higher chiral order of the $M1$ operator than what we consider here.

At other energies, our results agree with other theoretical results where available.~These include the results of Ref.~\cite{FinkExp} obtained with Paris~\cite{Paris1980} and Bonn~\cite{Bonn1986} potentials.~Our results also agree with the results of Refs.~\cite{Rupak, Ando2006} and Ref.~\cite{Richardson2025} at $E=0.5, 1$ MeV and $E$ ranging from $0.1$ to $0.5$ MeV, respectively as well as with the results of Ref.~\cite{Acharya2022}.~Reference~\cite{Acharya2022} employs the semilocal momentum-space regularized chiral two-nucleon potentials up to N4LO~\cite{Reinert2018} together with electromagnetic currents including one-body plus two-body one-pion exchange electromagnetic currents up to N2LO derived in Ref.~\cite{SPastore2008} within the time-ordered perturbation theory presented in Ref.~\cite{Piarulli2013}, and using the parameterization of Ref.~\cite{Borah2020}~to calculate the total capture cross section up to $E=1$~MeV.  Since Ref.~\cite{Acharya2022} does not include N3LO terms for the $M1$ transition operator, the authors obtain the same capture cross section result at $E=1.2625 \cdot 10^{-8}$~MeV as in our work.


We note that the capture cross section at low energies is known to be proportional to $1/k$.~We plot the $1/k$ dependence of the cross section in Fig.~\ref{Fig1InsetCAPCSInset} by fitting the lowest energy experimental data of Ref.~\cite{CoxExp} and find that the $1/k$ approximation works well up to $E=0.01$ MeV.

We compare our photodisintegration cross section results with experimental data~\cite{BarnesExp, SkopikExp,AhrensExp,BirenbaumExp, BernabeiExp, MorehExp, GraeveExp, HaraExp,ChenExp}.~Our results agree well with experimental data except with those of Ref.~\cite{GraeveExp} at photon energies $E_{\gamma}=16, 18$, and \mbox{$19$ MeV}, where the discrepancies are at the order of a few percent and with Ref.~\cite{BernabeiExp} at $E_{\gamma}=14.7$ MeV, where we also have a similar discrepancy.  However, we agree at $E_{\gamma}=18$ and $19$ MeV with the results of Ref.~\cite{SkopikExp}.

Generally, our results agree well with the theoretical calculations of Refs.~\cite{BirenbaumExp, BernabeiExp, Partovi1964,ChenSavage}.  In particular, our results agree with the theoretical results of Refs.~\cite{BirenbaumExp,Partovi1964}, which use the Hamada-Johnston potential \cite{HamadaJohnston1962}, except for a slight discrepancy with the result at $E_{\gamma}=20$ MeV at the order of a fraction of a percent.~As for the comparison with Ref.~\cite{BernabeiExp}, which uses an improved version \cite{Cambi1982} of the theory of Ref.~\cite{Partovi1964}, we obtain a few percent discrepancy at \mbox{$E_{\gamma}=14.7$} MeV and agree with the theory at \mbox{$E_{\gamma}=19.3$ MeV}.  Our results are also in agreement with those of Ref.~\cite{ChenSavage}, which uses pionless EFT.

\section{Conclusions}
\label{sec3}

\hspace{1.5mm }  As an initial application of the Efros method \cite{VDEfros} in our adaptation \cite{MethodsPaper2024}, we have calculated the reactions~\mbox{$np \rightarrow d\gamma$} and $ d\gamma \rightarrow np$ using the same modern inter-nucleon interaction and $M1$ operator employed in our previous work~\cite{DuWeijieHORSE}.  In particular, we used the LENPIC N4LO interaction~\cite{Epelbaum2015} as well as the $M1$ operator from the same $\chi$EFT up to N2LO~\cite{SPal2023}.~We used only the naive $E1$ operator since the NLO and N2LO corrections for $E1$ were shown to be negligible~\cite{SSarkerThesis}.~We have extended the results of Ref.~\cite{DuWeijieHORSE} by calculating the capture and photodisintegration reactions over broad energy ranges, namely, over $E$ ranging from astrophysical energies to $17.78$ MeV and over respective $E_{\gamma}$ up to $20$ MeV.~We have calculated the reactions at energies where, to our knowledge, no experimental or theoretical results are available, and, in some cases, neither are reported.~Working with the oscillator basis, we constructed scattering wave functions for the $np$ system by using infinite series in oscillator functions.

For the two-body problems, our approach can be compared with the HORSE method that requires the calculation of all two-body eigenstates.  For all quantities, we obtain accurate results from our adaptation of the Efros method using a limited number of SRFs as well as fewer oscillator quanta in the interaction than those needed in the HORSE method in Ref.~\cite{DuWeijieHORSE}.~We also find that our results generally compare well with experiment and other theoretical approaches.  

The situation is different for many-body applications in terms of computational feasibility.  In this case, our approach remains feasible because it obtains converged results with a limited number of SRFs accessible using modern NCSM codes.  This is not the case for the HORSE method which requires the computation of all eigenstates with definite $J$ separated from the center-of-mass motion that makes it impossible to utilize the HORSE method in combination with NCSM in many-body applications.

We utilized $S$-matrix pole location to obtain an accurate deuteron bound state.~In general, the method of $S$-matrix pole location is useful for accurately calculating bound states and resonances as was demonstrated in Ref.~\cite{MethodsPaper2024}.~We note, however, that we use a large basis to get accurate results for the deuteron.~This is because the deuteron is a weakly bound system and we have used a large $\hbar \Omega$ value.~It is also needed to get an accurate asymptotic behavior of the deuteron wave function because we include astrophysical energies in our calculations, where this asymptotic tail is important.

By obtaining accurate results with reduced interaction oscillator quanta and limited number of SRFs, we have demonstrated the applicability of the Efros method.  In combination with the NCSM, the Efros method will be promising for \textit{ab initio} studies of elastic scattering and reactions of light nuclei.~In addition, the Efros method will enable us to obtain converged and accurate predictions of various nuclear observables such as rms radii, quadrupole moments, asymptotic normalization constants, and electromagnetic transition rates.

 \section*{\NoCaseChange{ACKNOWLEDGEMENTS}}

We are thankful to J. P. Vary, P. Maris, V. D. Efros, E. Epelbaum, T. R. Richardson, and I. C. Reis for useful discussions. A. M. Shirokov is thankful for the hospitality of colleagues from Pacific National University (Khabarovsk, Russia) where a part of this work was done.  This work is supported by new faculty startup funding by the Institute of Modern Physics, Chinese Academy of Sciences, grant No.~E539951SBH, by the Gansu International Collaboration and Talents Recruitment Base of Particle Physics (2023–2027), by the Senior Scientist Program funded by Gansu Province grant No.~25RCKA008, by the Gansu ``Special Program for the Recruitment of Foreign Experts" under grant No.~26RCKA016, and by the Ministry of Science and Higher Education of the Russian Federation, project No.~FEME-2024-0005).

\section*{\NoCaseChange{DATA AVAILABILITY}}

The data that support the findings of this article are not publicly available. The data are available from the authors upon reasonable request.

\appendix
\renewcommand{\appendixname}{\ APPENDIX}

\section{ \ SCATTERING WAVE FUNCTION}
\label{AppendixA}

\hspace{1.5mm} For the convenience of the readers, we present here definitions of some functions utilized in our approach. All these definitions and additional details of the approach can be found in Ref.~\cite{MethodsPaper2024}.

The long-range behavior of the scattering wave 
functions is described by the functions
\begin{gather}
\tag{A1}
\eta^\pm_i(\vec{r}) 
=\tilde{h}^{\pm}_{l_i}(kr)\mathscr{Y}_{J M_{J}}^{l_{i} S_{i}}(\Omega_{\hat{r}}).
\end{gather}
Here,
\begin{gather}
\tag{A2}
\tilde{h}^{\pm}_{l}(kr)=\tilde{n}_{l}(kr) \pm ij_{l}(kr), 
\end{gather}
where $j_{l}(z)$~is the spherical Bessel function and 
the functions~$\tilde{n}_{l}(z)$
and~$\tilde{h}^{\pm}_{l}(z)$ behave asymptotically as
\begin{align}
\tilde{n}_{l}(z) &\mathop{\longrightarrow}_{z\to\infty} -n_{l}(z) 
\mathop{\longrightarrow}_{z\to\infty} \frac{1}{z}\cos\!\Big(z-\frac12\pi l\Big),
\tag{A3} \\
\tilde{h}^{\pm}_{l}(z)&\mathop{\longrightarrow}_{z\to\infty} { \pm ih^{(1,2)}_l(z)}
\mathop{\longrightarrow}_{z\to\infty}  \frac{i^{\mp l}}{z}e^{\pm iz}.
\tag{A4}
\end{align}
The singularity at the origin of the spherical Neumann~$n_{l}(z)$ and Hankel~$h^{(1,2)}_l(z)$ functions~is regularized in the functions~$\tilde{n}_{l}(z)$
and~$\tilde{h}^{\pm}_{l}(z)$ that makes it possible to avoid divergences of integrals calculated with them. 

$j_{l}(kr)$, $\tilde{n}_{l}(kr)$, and~$\tilde{h}^{\pm}_{l}(kr)$ can be expanded as
\begin{align}
\tag{A5}
j_{l}(kr)=&\sum_{n=0}^{\infty}S_{nl}(k)R_{nl}(r),\\
\tag{A6}
 \tilde{n}_l(kr)=&\sum_{n=0}^{\infty}C_{nl}(k)R_{nl}(r),\\
\tag{A7}
\tilde{h}^{\pm}_{l}(kr)=&\sum_{n=0}^{\infty}C^{\pm}_{n l}(k)R_{nl}(r),
\end{align}
where
\color{black}
\begin{align}
\label{eqS8}
\tag{A8}
S_{nl}(k)=&\:\sqrt{\frac{\pi b n!}{\Gamma(n+l+\frac{3}{2})}}(kb)^{l+1}e^{-\frac{k^{2}b^{2}}{2}}L_{n}^{l+\frac{1}{2}}(k^{2}b^{2}),
\\[1.5ex]
\label{eqS9}
C_{nl}(k)=&\:\frac{(-1)^{l}}{\Gamma(-l+\frac{1}{2})}\sqrt{\frac{\pi b n!}{\Gamma(n+l+\frac{3}{2})}}(kb)^{-l}e^{-\frac{k^{2}b^{2}}{2}} \notag \\
&\phantom{\Gamma(-l)} \times \tag{A9}
{_{1}F_{1}}\Big(\!-n-l-\frac{1}{2},-l+\frac{1}{2};k^{2}b^{2}\Big),
\\[1.5ex]
\color{black}
\label{eqS10}
\tag{A10}
\color{black} C^{\pm}_{n l}(k)=&\color{black}\frac{i}{2}[\:C_{n l}(k) \pm i S_{n l}(k)],
\end{align}
\color{black}
 $L_{n}^{\alpha}(z)$~is the associated Laguerre polynomial, ${_{1}F_{1}}(a,b;z)$ \color{black} is the confluent hypergeometric function, and 
\begin{equation}
\tag{A11}
R_{nl}(r)=(-1)^{n}\sqrt{\frac{2n!}{b^{3}\Gamma(n+l+\frac{3}{2})}}\Big(\frac{r}{b}\Big)^{l} e^{-\frac{r^{2}}{2b^{2}}}L_{n}^{l+\frac{1}{2}}\!\Big(\frac{r^{2}}{ b^{2}}\Big).
\end{equation}
Eqs.~\eqref{eqS8}-\eqref{eqS10} are the same as those in Ref.~\cite{MethodsPaper2024} up to an overall normalization.
\color{black}

\begin{figure}[t!]
\subfloat[]{%
  \includegraphics[width=1\columnwidth]{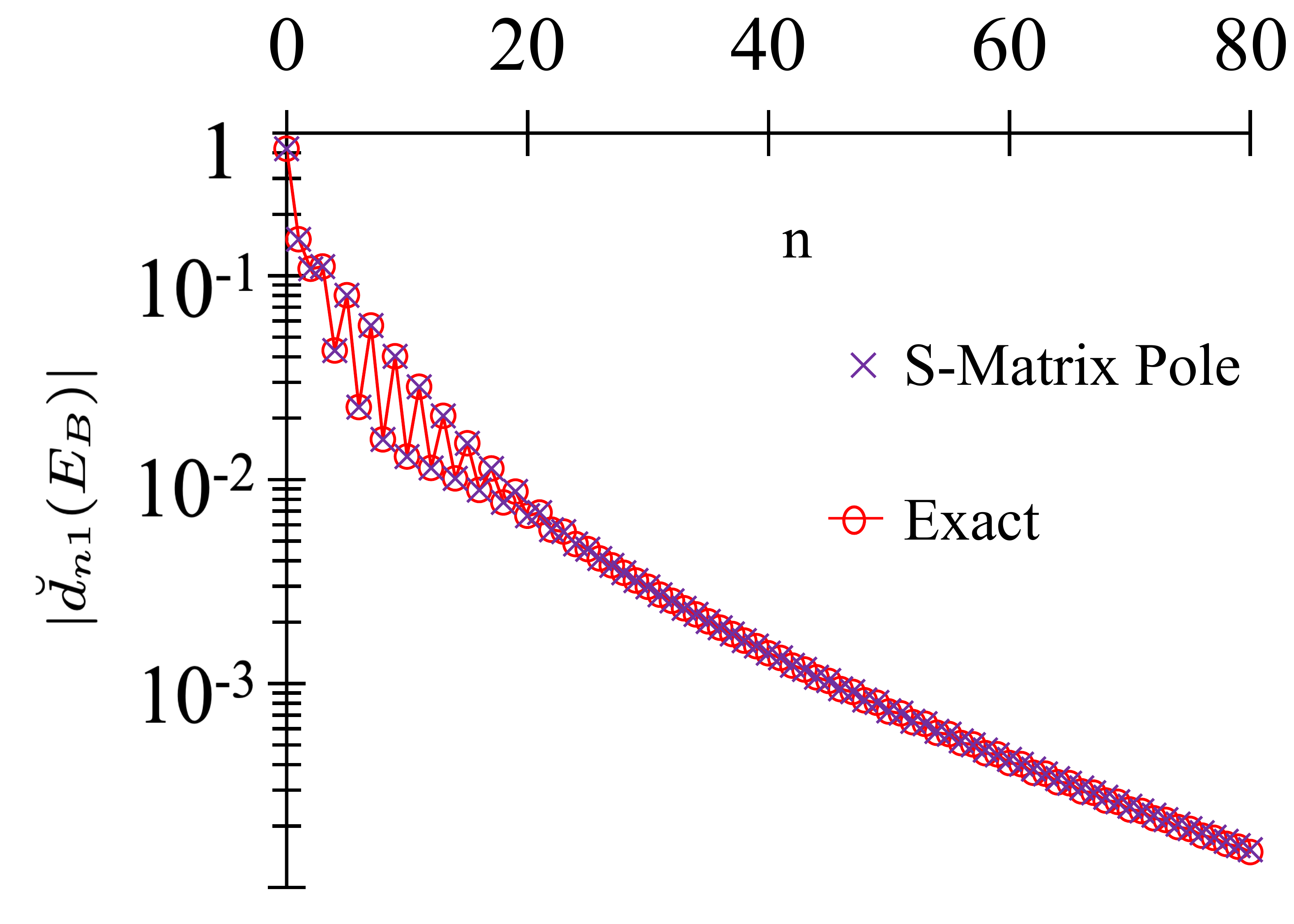}%
}\hfill
\subfloat[]{%
  \includegraphics[width=1\columnwidth]{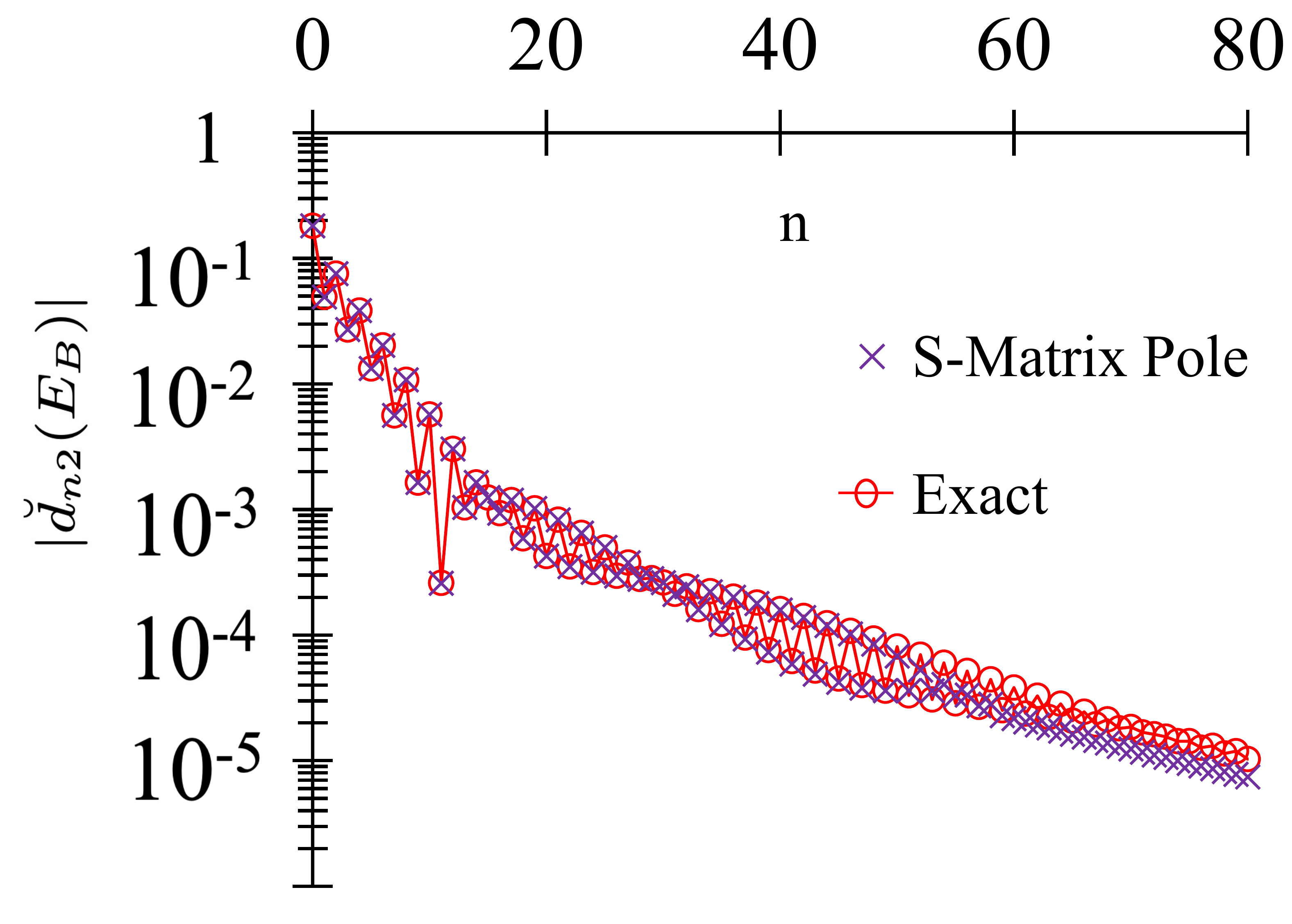}%
}\hfill
\caption{Absolute value of the a) $s$-wave bound state amplitude $|\breve{d}_{n1}(E_{B})|$ and b) $d$-wave bound state amplitude $|\breve{d}_{n2}(E_{B})|$ obtained via $S$-matrix pole location with $v=118$ at interaction truncation $N_{\textrm{max}}=120$ using the set of eigenfunction SRFs given in Eq.~\eqref{eq25} as compared with the exact result. \color{black}  
 }   
\label{figBS1}
\end{figure}

\section{\ BOUND STATE WAVE FUNCTION}
\label{AppendixB}


\hspace{1.5mm}  We discuss some features of the obtained deuteron wave function.  We expand the bound state wave function in an infinite series of oscillator functions $\phi_{ni}(\vec r)$,
\begin{equation}
\langle \vec r|\Psi^{J=1,S=1}\rangle=\sum_{i=1}^{2}\sum_{n=0}^{\infty}\breve{d}_{ni}(E_{B})\phi_{ni}(\vec r),
\label{amplitudeBound}
\tag{B1}
\end{equation}
where $\breve{d}_{ni}(E_{B})$ are the bound state expansion coefficients.  In Fig.~\ref{figBS1}, we present the absolute values of the bound state expansion coefficients $|\breve{d}_{ni}(E_{B})|$ in the oscillator basis obtained by the Efros method for radial quantum numbers ranging from $n=0$ up to $n=80$ and compare with the exact result.  For the $^3{S_{1}}$  partial wave, we see a high accuracy, including at large $n \geq 20$ beyond which the wave function exponentially decreases.  For the $^3{D_{1}}$ partial wave, the results are accurate only up to around $n=50$ beyond which our results start deviating from the exact ones.~At $n \geq 70$, that deviation is around 30\% from the exact result.  Note however, that since the tail of the $^3{D_{1}}$ partial wave at $n \geq 50$ has a negligible contribution to the bound state wave function (with the sum of amplitude squared for $n \geq 50$ being at the order of $10^{-8}$), this discrepancy at larger $n$ is unimportant for the purposes of computing the total capture cross section.  As in the case of bound state energy, the agreement with the exact result can be systematically improved by further increasing the interaction truncation $N_{\textrm{max}}$.

\vspace{5mm}
 
 \section{\ CROSS SECTIONS }
\label{AppendixC}

 \begin{table*}[t!]
\caption{Total capture cross section (in $\mu$b), $M1$ ($^1S_{0}$ partial wave) contribution to the total capture cross section, and $E1$ ($^3P_{0}$, $^3P_{1}$, and $^3PF_{2}$ partial waves) contribution to the total capture cross section at different energies $E$ (in MeV) using interaction truncations $N_{\textrm{max}}=50$ for $M1$ capture and $N_{\textrm{max}}=20$ for $E1$ capture.  Listed are $v$ values corresponding to converged calculations. }
\begin{tabular}{|c|c|c|c|c|c|c|c|c|c|c|c|c|}
\hline
$E$ & \multicolumn{2}{|c|}{$M1$ } & \multicolumn{6}{|c|}{$E1$ }  & Total & Other Theories & Experiment \\
\hline
 & $^1S_{0}$ & $v$ & $^3P_{0}$ & $v$  & $^3P_{1}$  & $v$ & $^3PF_{2}$ & $v$ & & &\\
 \hline
 \makecell{$1.2625 \cdot 10^{-8}$}  &  $3.191(43) \cdot 10^{5}$   & $15$ & - & - & - & - &\hspace{0.75mm}  - \hspace{0.25mm} & - &  $3.191(43) \cdot 10^{5}$    & \makecell{$3.34(3) \cdot 10^{5}$ \cite{TSPark1996}\\ $3.34(2) \cdot 10^{5}$ \cite{Ando2006} \\ \hspace{-1mm} $3.31(1) \cdot 10^{5}$ \cite{YHSong2009} \\ \hspace{-1mm} $3.35(5) \cdot 10^{5}$ \cite{Beane2015} \\ \hspace{-1mm} $3.31(2) \cdot 10^{5}$ \cite{Piarulli2013} \\ \hspace{-1mm} $3.21(1) \cdot 10^{5}$ \cite{Acharya2022}    } &  \makecell{$3.342(5)\cdot 10^{5}$ \hspace{-2 mm} \cite{CoxExp}\\ $3.326(7) \cdot 10^{5}$\cite{CokinosExp} \\ \hspace{-1.5 mm} $3.363_{-0.015}^{+0.012} \cdot 10^{5}$\cite{AndersonExp}   }\\
 \hline
 \makecell{$5 \cdot 10^{-7}$}   & \hspace{1mm} $5.1(1) \cdot 10^{4}$ \hspace{1mm} & $15$ & - & - & - & - & \hspace{0.75mm}  - \hspace{0.75mm} & - &   $5.1(1) \cdot 10^{4}$ & -
  & -\\
 \hline
 $5 \cdot 10^{-4}$  & \hspace{1mm} $1.59(2) \cdot 10^{3}$ \hspace{1mm} & $15$ & - & - & - & - & \hspace{0.75mm}  - \hspace{0.75mm}  & -&  $1.59(2) \cdot 10^{3} $ & \makecell{\hspace{-5mm} $1.67 \cdot 10^{3}$ \cite{Rupak} \\  \hspace{-1mm} $1.67(1) \cdot 10^{3}$ \cite{Ando2006}} & -\\
 \hline
 $1 \cdot 10^{-3}$  & \hspace{1mm} $1.12(2) \cdot 10^{3}$ \hspace{1mm} & $15$ & - & - & - & - &\hspace{0.75mm}  - \hspace{0.75mm} & -&  $1.12(2) \cdot 10^{3}$ & \makecell{\hspace{-5mm} $1.17 \cdot 10^{3}$ \cite{Rupak} \\  \hspace{-1mm} $1.17(1) \cdot 10^{3}$ \cite{Ando2006}} & -\\
 \hline
 $5 \cdot 10^{-3}$  & \hspace{1mm} $473(6)$ \hspace{1mm} & $15$ & - & - & - & - &\hspace{0.75mm}  - \hspace{0.75mm} & -& $473(6) $ & \makecell{\hspace{-3mm} $498 $ \cite{Rupak} \\ \hspace{1mm} $498(2) $ \cite{Ando2006}} & -\\
 \hline
 $0.01$  & \hspace{1mm} $313(3)$ \hspace{1mm} & $15$ & \hspace{0.75mm} $0.436(1)$ \hspace{0.75mm}  & $4$ & \hspace{0.75mm} $1.647(1)$ \hspace{0.75mm} & $5$ & \hspace{0.75mm} $2.53(2)$ \hspace{0.75mm}  & $3$ &  $318(3)$   &  \makecell{ \hspace{-5.5mm} $332 $ \cite{Rupak} \\ \hspace{-1mm}  $332(2) $ \cite{Ando2006}  } & $318(25)$ \cite{SuzukiExp} \\
 \hline
 $0.02$  & \hspace{1mm}  $197(2)$ \hspace{1mm} & 15 & \hspace{0.75mm} $0.613(2)$ \hspace{0.75mm}  & 4 & \hspace{0.75mm} 2.319(1) \hspace{0.75mm}  & 5 &  \hspace{0.75mm} 3.57(3) \hspace{0.75mm} & 3 &  $203(2) $   & - & $203(19)$ \cite{SuzukiExp} \\
 \hline
  $0.032$  & \hspace{0.75mm} $137(1)$ \hspace{0.75mm}  & $15$ & $0.771(2)$  & $4$  & $2.917(2)$ & $5$  & $4.49(3)$   & $3$ &    $146(1)$   & -& $151(7) $ \cite{SuzukiExp} \\
   \hline
   $0.04$  & \hspace{0.75mm} $116(2)$ \hspace{0.75mm}   & $13$  & \hspace{0.75mm} $0.859(2)$ \hspace{0.75mm}  & $4$  & \hspace{0.75mm} $3.250(2)$ \hspace{0.75mm} & $5$  & \hspace{0.75mm}  $5.00(4)$ \hspace{0.75mm}     & $3$ & $125(2)$    & - & -\\
 \hline
   $0.05$  & \hspace{0.75mm}  $95(1)$ \hspace{0.75mm}    & $13$  & \hspace{0.75mm} $0.956(3)$ \hspace{0.75mm}  & $4$  & \hspace{0.75mm} $3.618(2)$ \hspace{0.75mm} & $5$  & \hspace{0.75mm}  $5.57(4)$ \hspace{0.75mm}     & $3$ & $105(1) $   & \makecell{ \hspace{-3.5mm} $108 $ \cite{Rupak} \\ \hspace{1mm}  $108(1) $ \cite{Ando2006}  } & -\\
 \hline
 
    $0.05512$  & \hspace{0.75mm}  $86(1)$ \hspace{0.75mm}       & $13$ &  \hspace{0.75mm} $1.001(3)$ \hspace{0.75mm}    & $4$   &  \hspace{0.75mm} $3.790(2)$ \hspace{0.75mm}   & $5$   &  \hspace{0.75mm} $5.83(4)$ \hspace{0.75mm}       & $3$ & $97(1)$   & - &  \hspace{0.75mm} $89(4)$ \cite{TomyoExp}  \hspace{0.75mm} \\
 \hline
 
  $0.07395$  & \hspace{0.75mm}  $64(1)$ \hspace{0.75mm}       & $9$ &  \hspace{0.75mm} $1.149(3)$ \hspace{0.75mm}    & $4$   &   \hspace{0.75mm} $4.354(2)$ \hspace{0.75mm}   & $5$   &  \hspace{0.75mm} $6.70(6)$ \hspace{0.75mm}      & $3$ & $76(1)$  & - & \hspace{0.75mm} $76(4)$ \cite{TomyoExp} \hspace{0.75mm} \\
 \hline
 
   $0.08996$  &  \hspace{0.75mm}  $53.1(2)$ \hspace{0.75mm}       & $11$ &   \hspace{0.75mm} $1.257(3)$ \hspace{0.75mm}    & $4$   &  \hspace{0.75mm} $4.786(3)$ \hspace{0.75mm}  & $5$   & $7.34(6)$   & $3$ & \hspace{0.5mm} $66.5(2)$     & - & \hspace{0.75mm} $67(4)$ \cite{TomyoExp} \hspace{0.75mm} \\
 \hline
 
    $0.1$  & \hspace{0.75mm} $47.9(5)$ \hspace{0.75mm}   & $11$  & \hspace{0.75mm} $1.319(3)$ \hspace{0.75mm}  & $4$  & \hspace{0.75mm} $5.006(3)$ \hspace{0.75mm} & $5$  & \hspace{0.75mm}  $7.7(1)$ \hspace{0.75mm}     & $3$ & \hspace{0.5mm}  $62.0(6)$  & \makecell{ \hspace{2.5mm} $63.5(1) $ \cite{Rupak} \\ \hspace{2.5mm}  $63.4(3) $ \cite{Ando2006}  \\ \hspace{2.5mm}  $59(4) $ \cite{Richardson2025} } & -\\
 \hline
 
    $0.1722$  &  \hspace{0.75mm} $25.7(4)$ \hspace{0.75mm}     & $13$ &   \hspace{0.75mm} $1.674(4)$  \hspace{0.75mm}   & $4$   &  \hspace{0.75mm} $6.370(4)$ \hspace{0.75mm} & $5$   & $9.8(1)$     & $3$ & \hspace{0.5mm} $43.6(5)$   &  -   & \hspace{0.75mm} $44(9)$ \cite{TomyoExp} \hspace{0.75mm}  \\
 \hline
 
    $0.2$  & \hspace{0.75mm} $21.6(4)$ \hspace{0.75mm}   & $13$  & \hspace{0.75mm} $1.781(4)$ \hspace{0.75mm}  & $4$ & \hspace{0.75mm} $6.786(5)$ \hspace{0.75mm}  & $5$  & \hspace{0.75mm}  $10.5(2)$ \hspace{0.75mm}    & $3$ & \hspace{0.5mm}  $40.7(5)$     & \hspace{0.25mm} $40(2)$ \cite{Richardson2025} & -\\
 \hline
    $0.2397$  & \hspace{0.75mm} $17.7(1)$ \hspace{0.75mm}    & $15$  & \hspace{0.75mm} $1.915(5)$ \hspace{0.75mm} & $4$  & \hspace{0.75mm} $7.310(5)$ \hspace{0.75mm}   &  $5$  & \hspace{0.75mm}  $11.3(2)$ \hspace{0.75mm}    & $3$ & \hspace{0.5mm} $38.3(3)$  & \hspace{2.5mm} $37.9(1)$ \cite{Acharya2022} & -\\
 \hline
 
    $0.275$  & \hspace{0.75mm} $15.1(1)$ \hspace{0.75mm}     & $15$ &  \hspace{0.75mm} $2.019(5)$  \hspace{0.75mm}   & $4$   & \hspace{0.75mm} $7.718(6)$ \hspace{0.75mm} & $5$   & \hspace{0.75mm}  $12.0(2)$ \hspace{0.75mm}      & $3$ & \hspace{0.5mm} $36.8(3)$   & - & $35.2(2.4)$  \cite{NagaiExp} \hspace{0.75mm}   \\
 \hline
 
     $0.3$  & \hspace{0.75mm} $13.6(1)$ \hspace{0.75mm}    & $15$  & \hspace{0.75mm} $2.086(5)$ \hspace{0.75mm}  & $4$  & \hspace{0.75mm} $8.0(2)$ \hspace{0.75mm}    & $2$ &   \hspace{0.75mm}  $12.4(2)$ \hspace{0.75mm}  & $4$ &  \hspace{0.5mm} $36.0(4)$   & \hspace{0.25mm} $35(1)$ \cite{Richardson2025} & -\\
 \hline
      $0.3452$  & \hspace{0.75mm} $11.49(6)$ \hspace{0.75mm}    & $15$  &  \hspace{0.75mm} $2.194(5)$ \hspace{0.75mm}  & $4$  & \hspace{0.75mm} $8.4(2)$ \hspace{0.75mm}  & $2$ & \hspace{0.75mm}  $13.0(3)$ \hspace{0.75mm}      & $4$ & \hspace{0.5mm} $35.2(5)$   & \hspace{2.5mm} $34.6(1)$ \cite{Acharya2022} & -\\
 \hline
      $0.4$  & \hspace{0.75mm} $9.64(5)$ \hspace{0.75mm}    & $15$  & \hspace{0.75mm} $2.306(5)$ \hspace{0.75mm}  & $4$  & \hspace{0.75mm} $8.9(2)$ \hspace{0.75mm} & $2$ & \hspace{0.75mm}  $13.5(1)$ \hspace{0.75mm}     & $5$ &   \hspace{0.5mm} $34.3(3)$  & \hspace{0.25mm} $34(1)$ \cite{Richardson2025} & -\\
 \hline
      $0.5$  &  \hspace{0.75mm} $7.39(4)$ \hspace{0.75mm}     & $15$  & \hspace{0.75mm} $2.473(5)$ \hspace{0.75mm} & $4$  & \hspace{0.75mm} $9.6(2)$ \hspace{0.75mm} & $2$  &   \hspace{0.75mm}  $14.5(1)$ \hspace{0.75mm}  & $5$ & \hspace{0.5mm} $34.0(3)$    & \makecell{\hspace{2.5mm} $34.1(2) $ \cite{Rupak} \\ \hspace{2.5mm} $34.1(2) $ \cite{Ando2006} \\ \hspace{2.5mm} $33.4(5) $ \cite{Richardson2025}} & -\\
 \hline
      $0.6136$  & \hspace{0.75mm} $5.7(1)$ \hspace{0.75mm}  & $13$  & \hspace{0.75mm} $2.616(4)$ \hspace{0.75mm}   & $4$  & \hspace{0.75mm} $10.2(2)$ \hspace{0.75mm} & $2$ & \hspace{0.75mm}  $15.5(1)$ \hspace{0.75mm}   & $5$ &   \hspace{0.5mm} $34.0(4)$  &\hspace{2mm} $33.7(1)$ \cite{Acharya2022}  & -\\
 \hline
      $0.7766$  & \hspace{0.75mm} $4.4(1)$ \hspace{0.75mm}    & $13$  & \hspace{0.75mm}  $2.764(4)$ \hspace{0.75mm}   & $4$  & \hspace{0.75mm} $10.8(2)$ \hspace{0.75mm} & $2$  & \hspace{0.75mm}  $16.4(1)$ \hspace{0.75mm}     & $5$ &   \hspace{0.5mm} $34.4(3)$   & \hspace{2mm} $34.1(1)$ \cite{Acharya2022} & -\\
 \hline
      $1$  &   \hspace{0.75mm} $3.29(4)$  \hspace{0.75mm} & $13$  & \hspace{0.75mm} $2.891(3)$ \hspace{0.75mm}   & $4$  & \hspace{0.75mm} $11.5(1)$ \hspace{0.75mm} & $2$  & \hspace{0.75mm}  $17.3(1)$ \hspace{0.75mm}    & $5$ &   \hspace{0.5mm} $35.0(2)$ & \makecell{\hspace{2.5mm} $34.9(2) $ \cite{Rupak} \\ \hspace{2.5mm} $34.9(3) $ \cite{Ando2006} \\ \hspace{2.5mm} $34.1(1) $ \cite{Richardson2025}}  & -\\
 \hline
      $2$  &  \hspace{0.75mm} $1.65(2)$  \hspace{0.75mm}   & $9$  & \hspace{0.75mm} $2.998(3)$ \hspace{0.75mm}  & $2$  & \hspace{0.75mm} $12.47(1)$ \hspace{0.75mm} & $2$  & \hspace{0.75mm}  $18.56(3)$ \hspace{0.75mm}    & $3$ &  \hspace{0.5mm} $35.7(1)$   & \hspace{2.5mm} $34.3(3) $ \cite{Richardson2025} & -\\
 \hline
      $3$  &  \hspace{0.75mm} $1.13(1)$ \hspace{0.75mm}   & $13$  & \hspace{0.75mm} $2.84(4)$ \hspace{0.75mm}  & $2$  & \hspace{0.75mm} $12.46(5)$ \hspace{0.75mm} & $2$  & \hspace{0.75mm}  $18.4(2)$ \hspace{0.75mm}     & $3$ &  \hspace{0.5mm} $34.9(2)$  & \hspace{2.5mm} $33.1(4) $ \cite{Richardson2025} & -\\
 \hline
      $4$  &  \hspace{0.75mm} $0.89(1)$ \hspace{0.75mm}    & $13$  & \hspace{0.75mm} $2.62(5)$ \hspace{0.75mm}  & $2$  & \hspace{0.75mm} $12.2(1)$ \hspace{0.75mm} & $2$  & \hspace{0.75mm}  $17.9(2)$ \hspace{0.75mm}     & $3$  &  \hspace{0.5mm} $33.6(4)$ &  \hspace{2.5mm} $31.7(4) $ \cite{Richardson2025} & -\\
 \hline
      $5$  & \hspace{0.75mm} $0.76(1)$ \hspace{0.75mm}  & $9$   & \hspace{0.75mm} $2.43(2)$ \hspace{0.75mm}    & $4$ & \hspace{0.75mm} $11.9(1)$ \hspace{0.75mm}  & $2$  & \hspace{0.75mm}  $17.1(3)$ \hspace{0.75mm}  & $3$  &   \hspace{0.5mm} $32.2(4)$ &  \hspace{2.5mm} $30.2(4) $ \cite{Richardson2025}  & -\\
 \hline
      $6$  &  \hspace{0.75mm} $0.65(1)$ \hspace{0.75mm}   & $9$  & \hspace{0.75mm} $2.23(3)$ \hspace{0.75mm}   & $4$  & \hspace{0.75mm} $11.5(2)$ \hspace{0.75mm} & $2$  & \hspace{0.75mm}  $16.4(3)$ \hspace{0.75mm}     & $3$ &  \hspace{0.5mm} $30.8(5)$   &  \hspace{2.5mm} $28.9(3) $ \cite{Richardson2025}  & -\\
 \hline
      $7.2$  &  \hspace{0.75mm} $0.584(4)$ \hspace{0.75mm}    & $11$   &  \hspace{0.75mm} $1.93(4)$ \hspace{0.75mm}    & $3$  & \hspace{0.75mm} $11.2(2)$ \hspace{0.75mm} & $2$ &  \hspace{0.75mm}  $15.5(3)$ \hspace{0.75mm}    & $3$ &  \hspace{0.5mm} $29.2(5)$    &  \hspace{2.5mm} $27.5(3) $ \cite{Richardson2025}  & $30.6(1.8) $ \cite{GhemoExp}  \\
 \hline
      $8$  & \hspace{0.75mm} $0.547(6)$ \hspace{0.75mm}  & $11$  & \hspace{0.75mm} $1.80(3)$ \hspace{0.75mm}   & $3$  & \hspace{0.75mm} $10.9(2)$ \hspace{0.75mm} & $2$ & \hspace{0.75mm}  $14.9(3)$ \hspace{0.75mm}    & $6$  &  \hspace{0.5mm} $28.2(5)$     &  \hspace{2.5mm} $26.6(3) $ \cite{Richardson2025} & -\\
 \hline
      $9$  & \hspace{0.75mm} $0.51(1)$ \hspace{0.75mm}   & $11$  & \hspace{0.75mm} $1.65(2)$ \hspace{0.75mm}  & $3$  & \hspace{0.75mm} $10.8(2)$ \hspace{0.75mm}   & $3$  & \hspace{0.75mm}  $14.3(2)$ \hspace{0.75mm}     & $3$ &  \hspace{0.5mm} $27.3(4)$      &  \hspace{2.5mm} $25.7(2) $ \cite{Richardson2025} & -\\
 \hline
    $9.65$  & \hspace{0.75mm} $0.49(1)$ \hspace{0.75mm}   & $11$ & \hspace{0.75mm} $1.57(2)$ \hspace{0.75mm}  & $3$  & \hspace{0.75mm} $10.7(2)$ \hspace{0.75mm}   & $3$  &   \hspace{0.75mm} $13.9(2)$ \hspace{0.75mm}  & $3$  & \hspace{0.5mm} $26.7(4)$      &  \makecell{\hspace{0.5mm} $ 27.0$~\cite[Paris]{FinkExp} \\  \hspace{0.5mm} $26.8$~\cite[Bonn]{FinkExp}} & -\\
 \hline
      $10$  & \hspace{0.75mm} $0.474(6)$ \hspace{0.75mm}   & $11$  & \hspace{0.75mm} $1.52(2)$ \hspace{0.75mm}  & $3$  & \hspace{0.75mm} $10.7(2)$ \hspace{0.75mm}  & $3$  & \hspace{0.75mm}  $13.7(2)$ \hspace{0.75mm}     & $3$   & \hspace{0.5mm} $26.4(4)$       &  \hspace{2.5mm} $24.8(2) $ \cite{Richardson2025} & -\\
 \hline
            $11.1$  & \hspace{0.75mm} $0.441(4)$ \hspace{0.75mm}   & 11  &  \hspace{0.75mm} $1.39(1)$ \hspace{0.75mm}   & $3$ & \hspace{0.75mm}  $10.5(2)$ \hspace{0.75mm} & $3$ &  \hspace{0.75mm}  $13.1(1)$ \hspace{0.75mm}      & $3$  &  \hspace{0.5mm} $25.4(3)$     &  \makecell{\hspace{0.5mm} $ 25.7$~\cite[Paris]{FinkExp} \\  \hspace{0.5mm} $25.5 $~\cite[Bonn]{FinkExp} }   & -  \\
            \hline
                  $12.5$  & \hspace{0.75mm} $0.405(3)$ \hspace{0.75mm}    & $11$  & \hspace{0.75mm} $1.242(4)$ \hspace{0.75mm}    & $3$ &  \hspace{0.75mm} $10.0(2)$ \hspace{0.75mm}    & $2$  &  \hspace{0.75mm} $12.36(6)$ \hspace{0.75mm}  & $3$  &  \hspace{0.5mm} $24.1(3)$       & \makecell{\hspace{0.5mm} $ 24.5$~\cite[Paris]{FinkExp} \\  \hspace{0.5mm} $24.4 $~\cite[Bonn]{FinkExp}}   & $26.6(1.5) $ \cite{StiehlerExp}   \\
                      \hline
                                        $12.8$  & \hspace{0.75mm} $0.398(2)$ \hspace{0.75mm}      & $11$  & \hspace{0.75mm} $1.213(3)$ \hspace{0.75mm}     & $3$ & \hspace{0.75mm} $10.0(2)$ \hspace{0.75mm}      & $2$  & \hspace{0.75mm} $12.2(1)$ \hspace{0.75mm}    & $3$  &  \hspace{0.5mm} $23.8(3)$         & \hspace{2.5mm} $22.7(1) $ \cite{Richardson2025}    & $24.2(7) $ \cite{MichelExp}     \\
                      \hline
                               $13.75$  & \hspace{0.75mm} $0.378(1)$ \hspace{0.75mm}  & $11$  &   \hspace{0.75mm} $1.13(1)$ \hspace{0.75mm}  &  $3$  & \hspace{0.75mm} $9.9(2)$ \hspace{0.75mm}  & $2$  &  \hspace{0.75mm} $11.7(1)$ \hspace{0.75mm}    & $3$  &  \hspace{0.5mm} $23.1(3)$      & \makecell{\hspace{0.5mm} $ 23.7$~\cite[Paris]{FinkExp} \\  \hspace{0.5mm} $23.5 $~\cite[Bonn]{FinkExp}}   & -   \\
                                   \hline
                                            $15$  & \hspace{0.75mm} $0.362(1)$ \hspace{0.75mm}   & $13$   &  \hspace{0.75mm} $1.02(1)$ \hspace{0.75mm}   & $3$ & \hspace{0.75mm} $10.0(2)$ \hspace{0.75mm}  & $4$ & \hspace{0.75mm} $11.2(2)$ \hspace{0.75mm}  & $3$  &    \hspace{0.5mm} $22.57(36)$        &  \hspace{2.5mm} $21.37(5) $ \cite{Richardson2025} & -\\
             \hline
              $17.78$  & \hspace{0.75mm} $0.325(1)$ \hspace{0.75mm}   & $13$   &  \hspace{0.75mm} $0.825(5)$  \hspace{0.75mm}   & $5$  & \hspace{0.75mm} $9.4(2)$  \hspace{0.75mm}  & $5$  & \hspace{0.75mm} $10.2(1)$ \hspace{0.75mm}  & $10$  &  \hspace{0.5mm} $20.8(3)$       &  - & -\\
             \hline
 \end{tabular}
\label{tab5}
\end{table*}

 \begin{table*}[t!]
\caption{The same as Table~\ref{tab5} but for the total photodisintegration cross section.  }
\begin{tabular}{|c|c|c|c|c|c|c|c|c|c|c|c|c|}
\hline
$E_{\gamma}$ & \multicolumn{2}{|c|}{$M1$ } & \multicolumn{6}{|c|}{$E1$ } & Total & Other Theories & Experiment \\
\hline
 & $^1S_{0}$ & $v$ & $^3P_{0}$ & $v$  & $^3P_{1}$  & $v$ & $^3PF_{2}$ & $v$ & &  &\\
 \hline
   $2.233$ & \hspace{-0.5mm} $393(4)$ \hspace{0.5mm}    & $15$  & -   & -  & -     & - &  -    & - &  \hspace{0.5mm} $393(4)$ \hspace{0.5mm}  & - & - \\
  \hline
    $2.273$ & \hspace{-0.5mm} $573(7)$ \hspace{0.5mm}    & $13$  &  \hspace{0.5mm} $5.79(2)$ \hspace{0.5mm}  & $4$   & \hspace{0.5mm} $21.91(1)$ \hspace{0.5mm}      & $5$   &  \hspace{0.5mm} $33.7(3)$ \hspace{0.5mm}     & $3$ & \hspace{0.5mm} $634(7) $ \hspace{0.5mm}    & \hspace{0.5mm} - \hspace{0.5mm}    & - \\
  \hline
  $2.33$ & \hspace{-0.5mm} $553(8)$ \hspace{0.5mm}    & $11$  &  \hspace{0.5mm} $16.74(4)$ \hspace{0.5mm}  & $4$   & \hspace{0.5mm} $63.51(4)$ \hspace{0.5mm}      & $5$   &   $98(1)$ \hspace{0.5mm}     & $3$ & \hspace{0.5mm} $731(9) $ \hspace{0.5mm}    & \hspace{0.5mm} - \hspace{0.5mm}    & \makecell{   \hspace{-1mm} $683(95)$  \cite{HaraExp} \hspace{-1mm}\\    \hspace{-1mm} $671(103)$ \cite{ChenExp}  } \\
  \hline
      $2.433$ & \hspace{-0.5mm}  $453(9)$   \hspace{0.5mm}    & $13$  &  \hspace{0.5mm} $40.3(1)$ \hspace{0.5mm}  &   $4$ &  \hspace{0.5mm} $153.6(1)$   \hspace{0.5mm}      & $5$  &  \hspace{0.5mm}  $237(3)$ \hspace{0.5mm}     & $3$  & \hspace{0.5mm} 884(12) \hspace{0.5mm}    & \hspace{0.5mm} - \hspace{0.5mm}    & \hspace{-1mm} $856(61)$ \cite{ChenExp} \hspace{0.5mm}    \\
        \hline
    $2.52$ &  \hspace{-0.5mm}    $403(2)$ \hspace{0.5mm}    & $15$  &  \hspace{0.5mm} $60.8(2)$ \hspace{0.5mm}      & $4$   &  \hspace{0.5mm} $233(5)$ \hspace{0.5mm}     & $2$   &  \hspace{0.5mm} $361(7)$ \hspace{0.5mm}  & $3$ &  \hspace{-3.5mm} $1.06(1) \cdot 10^{3}$ \hspace{0.5mm}    & -    & \hspace{-1mm} $983(10)$ \cite{HaraExp} \hspace{0.5mm}  \\
    \hline
 $2.62$ &  \hspace{-0.5mm}  $352(2)$ \hspace{0.5mm}   & $15$  &  \hspace{0.5mm} $83.2(2)$ \hspace{0.5mm}  & $4$  & \hspace{0.5mm} $321(6)$ \hspace{0.5mm}  & $2$  & \hspace{0.5mm} $488(3)$ \hspace{0.5mm}   & $5$ & \hspace{-3.5mm} $1.24(1) \cdot 10^{3}$ \hspace{0.5mm}    & \hspace{0.5mm} $1.25(5) \cdot 10^{3}$ \cite{ChenSavage} \hspace{0.5mm}     & - \\
  \hline
   $2.754$    &  \hspace{-0.5mm}   $302(2)$    \hspace{0.5mm}   & $15$    &  \hspace{0.5mm} $110.2(2)$   \hspace{0.5mm}  & $4$  & \hspace{0.5mm}  $427(7)$ \hspace{0.5mm}  & $2$ & \hspace{0.5mm} $649(4)$ \hspace{0.5mm}   &  $5$  & \makecell{ $1.488(12) \cdot 10^{3}$  }      &  -  & \hspace{-3.5mm} $1.456(45)  \cdot 10^{3}$ \cite{MorehExp} \hspace{0.5mm} \\
  \hline
 $2.76$    &  \hspace{-0.5mm}    $299(2)$  \hspace{0.5mm}   & $15$   &  \hspace{0.5mm} $111.3(2)$  \hspace{0.5mm}  & $4$  & \hspace{0.5mm} $432(7)$ \hspace{0.5mm}  & $2$  & \hspace{0.5mm} $656(4)$ \hspace{0.5mm}   & $5$ &   \hspace{-3.5mm} $1.50(1) \cdot 10^{3}$ \hspace{0.5mm}  &  $1.50(6) \cdot 10^{3}$ \cite{ChenSavage}    & -\\
  \hline
    $2.79$ &  \hspace{-0.5mm}   $290(1)$  \hspace{0.5mm}    & $15$  &  \hspace{0.5mm} $116.8(2)$  \hspace{0.5mm}      & $4$   &  \hspace{0.5mm} $453(7)$  \hspace{0.5mm}     &    $2$ &  \hspace{0.5mm} $689(4)$  \hspace{0.5mm}    & $5$  & \hspace{-4mm}  $1.55(1) \cdot 10^{3}$   &  -    & \hspace{-1mm} $1.47(12)  \cdot 10^{3}$ \cite{HaraExp} \hspace{0.5mm}  \\
    \hline
        $2.906$ &  \hspace{-0.5mm}   $256(5)$    \hspace{0.5mm}    & $13$  &  \hspace{0.5mm}  $135.9(2)$ \hspace{0.5mm}      &  $4$  &  \hspace{0.5mm} $531(7)$  \hspace{0.5mm}     &  $2$   &  \hspace{0.5mm} $805(5)$    \hspace{0.5mm}    & $5$   &   \makecell{ $1.727(17) \cdot 10^{3}$  }       &  -    & \hspace{-1mm} $1.712(102)  \cdot 10^{3}$ \cite{ChenExp} \hspace{0.5mm}  \\
    \hline
      $3.23$ &  \hspace{-0.5mm}   $197(2)$  \hspace{0.5mm}   & $13$  & \hspace{0.5mm} $174.8(1)$  \hspace{0.5mm}     & $4$   & \hspace{0.5mm} $693(7)$  \hspace{0.5mm}    & $2$  & \hspace{0.5mm} $1.047(1) \cdot 10^{3}$  \hspace{0.5mm}    & $5$ & \hspace{-3.5mm} $2.11(2) \cdot 10^{3}$  \hspace{0.5mm}   & -    & \hspace{-1mm} $2.04(17)  \cdot 10^{3}$ \cite{HaraExp} \hspace{0.5mm}  \\
      \hline
          $3.436$ &  \hspace{-0.5mm} $177(4)$   \hspace{0.5mm}    & $12$   &  \hspace{0.5mm} $190.3(2)$   \hspace{0.5mm}      & $4$    &  \hspace{0.5mm} $762(6)$  \hspace{0.5mm}     &   $2$  &  \hspace{0.5mm} $1.134(15) \cdot 10^{3}$    \hspace{0.5mm}    & $3$  &  \makecell{ $2.263(25) \cdot 10^{3}$  }       &  -    & \hspace{-1mm} $2.222(137)  \cdot 10^{3}$ \cite{ChenExp} \hspace{0.5mm}  \\
    \hline
        $3.69$ &  \hspace{-0.5mm}    $150(1)$  \hspace{0.5mm}     & $11$  &  \hspace{0.5mm} $205(2)$  \hspace{0.5mm}  & $3$   &  \hspace{0.5mm} $820(4)$  \hspace{0.5mm}     & $2$   & \hspace{0.5mm} $1.222(9) \cdot 10^{3}$  \hspace{0.5mm}     &    $3$   &  \makecell{ \hspace{-1.5mm} $2.397(17) \cdot 10^{3}$  }      &   -    & \makecell{   \hspace{-1mm} $2.29(18)  \cdot 10^{3}$ \cite{HaraExp}\hspace{0.5mm}\\    \hspace{-1mm} $2.373(143)  \cdot 10^{3}$ \cite{ChenExp}  } \hspace{0.5mm} \\
        \hline
      $3.859$ &  \hspace{-0.5mm}  $135(2)$  \hspace{0.5mm}    & $7$  &  \hspace{0.5mm}  $206.8(1)$ \hspace{0.5mm}      &  $4$  &  \hspace{0.5mm} $846(3)$  \hspace{0.5mm}     &  $2$   &  \hspace{0.5mm} $1.260(6) \cdot 10^{3}$  \hspace{0.5mm}    & $3$  &  \hspace{-0.2mm} $2.448(11) \cdot 10^{3}$  \hspace{0.5mm}    &  -    & \hspace{-1mm} $2.464(145)  \cdot 10^{3}$ \cite{ChenExp} \hspace{0.5mm}  \\
    \hline
          $4.181$ &  \hspace{-0.5mm}  $118(2)$  \hspace{0.5mm}    & $9$  &  \hspace{0.5mm}  $210.4(2)$ \hspace{0.5mm}      &  $2$  &  \hspace{0.5mm} $873(1)$   \hspace{0.5mm}     &  $2$   &  \hspace{0.5mm}  $1.300(2) \cdot 10^{3}$  \hspace{0.5mm}    & $3$   & \hspace{-1mm} $2.501(5) \cdot 10^{3}$  \hspace{0.5mm}    &  -    & \hspace{-1mm} $2.532(148)  \cdot 10^{3}$ \cite{ChenExp} \hspace{0.5mm}  \\
    \hline
   $4.45$  &  \hspace{-0.5mm}  $101(2)$  \hspace{0.5mm}    & $11$  &  \hspace{0.5mm} $209(1)$  \hspace{0.5mm}  & $2$  & \hspace{0.5mm} $881(1)$ \hspace{0.5mm}  & $2$  &\hspace{0.5mm} $1.310(2) \cdot 10^{3}$ \hspace{0.5mm}   & $3$ & \hspace{0.5mm} $2.50(1) \cdot 10^{3} $ \hspace{0.5mm}     &   $2.48(9) \cdot 10^{3}$ \cite{ChenSavage}    & \hspace{-3mm} $2.43(17)  \cdot 10^{3}$ \cite{BarnesExp} \hspace{0.5mm} \\
  \hline
    $4.53$ &  \hspace{-0.5mm}   $100(1)$  \hspace{0.5mm}   & $13$ &  \hspace{0.5mm} $208(1)$  \hspace{0.5mm}     & $2$  &  \hspace{0.5mm} $881(1)$  \hspace{0.5mm}   & $2$   &  \hspace{0.5mm} $1.309(3) \cdot 10^{3}$  \hspace{0.5mm}   & $3$ &  \hspace{0.5mm} $2.50(1) \cdot 10^{3}$  \hspace{0.5mm}   & -    & \hspace{-1mm} $2.48(19)  \cdot 10^{3}$ \cite{HaraExp} \hspace{0.5mm} \\
    \hline
        $4.58$ & \hspace{0.5mm}   $98.5(6)$  \hspace{0.5mm}   & $13$  &   \hspace{0.5mm} $208(1)$  \hspace{0.5mm}   & $2$  &  \hspace{0.5mm} $881(1)$  \hspace{0.5mm}    &   $2$ &  \hspace{0.5mm} $1.309(4) \cdot 10^{3}$  \hspace{0.5mm}   & $3$ &  \hspace{0.5mm} $2.50(1) \cdot 10^{3}$  \hspace{0.5mm}   &   -   & \hspace{-1mm} $2.41(17)  \cdot 10^{3}$ \cite{HaraExp} \hspace{0.5mm} \\
    \hline
    $5.97$ &  \hspace{-1.5mm}  $62.0(5)$  & $13$  & \hspace{0.5mm} $176(3)$  \hspace{0.5mm}  & $2$ &  \hspace{0.5mm} $808(5)$ \hspace{0.5mm}  & $2$  & \hspace{0.5mm} $1.186(13) \cdot 10^{3}$ \hspace{0.5mm}  & $3$ & \makecell{ $2.232(22) \cdot 10^{3}$  }     &  \makecell{  $2.2 \cdot 10^{3}$~\cite{BirenbaumExp}  \\ \hspace{-1mm}   $2.21(8) \cdot 10^{3}$ \cite{ChenSavage}  }      & \hspace{1.5mm} $2.162(99)  \cdot 10^{3}$ \cite{BirenbaumExp} \hspace{0.5mm} \\
  \hline
   $6.14$  & \hspace{0.5mm}  $59.2(5)$  \hspace{0.5mm}   & $13$ &  \hspace{0.5mm} $171(3)$  \hspace{0.5mm}  & $2$  & \hspace{0.5mm} $795(6)$ \hspace{0.5mm}   & $2$  &\hspace{0.5mm} $1.165(14) \cdot 10^{3}$ \hspace{0.5mm}  & $3$ & \hspace{0.5mm} $2.19(2) \cdot 10^{3}$ \hspace{0.5mm}   &   $2.17(8) \cdot 10^{3}$ \cite{ChenSavage}    & \hspace{-2.5mm} $2.19(10)  \cdot 10^{3}$ \cite{BarnesExp} \hspace{0.5mm} \\
       \hline
       $6.632$ &  \hspace{0.5mm}   $52.0(3)$  \hspace{0.5mm}    & $13$   &  \hspace{0.5mm}  $156(3)$ \hspace{0.5mm}      & $3$   &  \hspace{0.5mm} $757(6)$  \hspace{0.5mm}     &   $2$  &  \hspace{0.5mm} $ 1.103(15) \cdot 10^{3}$   \hspace{0.5mm}    & $3$  & \makecell{ $2.068(25) \cdot 10^{3}$  }       &  -    & \hspace{-1mm} $2.203(133)  \cdot 10^{3}$ \cite{ChenExp} \hspace{0.5mm}  \\
  \hline
         $7.089$ &  \hspace{0.5mm} $45(1)$   \hspace{0.5mm}    &  $12$ &  \hspace{0.5mm} $149(1)$  \hspace{0.5mm}      &  $4$  &  \hspace{0.5mm} $722(7)$  \hspace{0.5mm}     &   $2$  &  \hspace{0.5mm}  $1.045(16) \cdot 10^{3}$ \hspace{0.5mm}    &  $3$ &    \makecell{ $1.961(24) \cdot 10^{3}$  }     &  -    & \hspace{-1mm} $2.030(124)  \cdot 10^{3}$ \cite{ChenExp} \hspace{0.5mm}  \\
  \hline
 
   $7.25$  & \hspace{0.5mm}   $45(1)$  \hspace{0.5mm}   & $9$  &  \hspace{0.5mm} $145(1)$  \hspace{0.5mm}   & $4$  & \hspace{0.5mm} $710(7)$ \hspace{0.5mm}    & $2$  & \hspace{0.5mm} $1.025(16) \cdot 10^{3}$ \hspace{0.5mm}  & $3$ &   \makecell{ $1.925(25) \cdot 10^{3}$  }      &\makecell{ $1.94 \cdot 10^{3}$~\cite{BirenbaumExp} \\ \hspace{-1mm}   $1.90(7) \cdot 10^{3}$ \cite{ChenSavage}  }     & \hspace{-3.5mm} $1.882(11)  \cdot 10^{3}$ \cite{BirenbaumExp} \hspace{0.5mm} \\
  \hline
  $7.39$   & \hspace{0.5mm}  $44(1)$  \hspace{0.5mm}   & $9$  & \hspace{0.5mm} $142(1)$  \hspace{0.5mm}    & $4$  &  \hspace{0.5mm} $699(7)$  \hspace{0.5mm}    & $2$  & \hspace{0.5mm} $1.008(16) \cdot 10^{3}$ \hspace{0.5mm}   & $3$ &  \hspace{0.5mm} $1.89(3) \cdot 10^{3}$ \hspace{0.5mm}   & $1.87(7) \cdot 10^{3}$ \cite{ChenSavage}      & \hspace{-2.5mm} $1.84(15)  \cdot 10^{3}$ \cite{BarnesExp} \hspace{0.5mm} \\
  \hline
  $7.6$   &  \hspace{-0.5mm}  $41.7(6)$  \hspace{0.5mm}  & $9$  &  \hspace{0.5mm} $137(1)$  \hspace{0.5mm}    & $4$  &  \hspace{0.5mm} $684(7)$  \hspace{0.5mm}   & $2$  & \hspace{0.5mm} $983(16)$ \hspace{0.5mm}   & $3$ &  \makecell{ $1.845(25) \cdot 10^{3}$  }      &  \makecell{  $1.87 \cdot 10^{3}$~\cite{BirenbaumExp} \\ \hspace{-1mm}   $1.83(7) \cdot 10^{3}$ \cite{ChenSavage}  }      & \hspace{-3.9mm} $1.803(16)  \cdot 10^{3}$ \cite{BirenbaumExp} \hspace{0.5mm}\\
  \hline
  $7.64$   &  \hspace{-0.5mm}  $41.3(6)$  \hspace{0.5mm}   & $9$  & \hspace{0.5mm} $136(1)$  \hspace{0.5mm}   & $4$  &  \hspace{0.5mm} $681(7)$  \hspace{0.5mm}  & $2$  & \hspace{0.5mm} $978(16)$ \hspace{0.5mm}   & $3$ &  \makecell{ $1.836(26) \cdot 10^{3}$  }     &  \makecell{   $1.86 \cdot 10^{3}$~\cite{BirenbaumExp} \\ \hspace{-1mm}   $1.82(7) \cdot 10^{3}$ \cite{ChenSavage}  }   & \hspace{-3.5mm} $1.810(28)  \cdot 10^{3}$ \cite{BirenbaumExp} \hspace{0.5mm} \\
  \hline
   $8.14$  & \hspace{-0.5mm}   $36.8(3)$  \hspace{0.5mm}   & $9$  & \hspace{0.5mm} $125(1)$  \hspace{0.5mm}    & $4$  &  \hspace{0.5mm} $646(8)$  \hspace{0.5mm}   & $2$  & \hspace{0.5mm} $920(17)$ \hspace{0.5mm}   & $3$ & \hspace{0.5mm} $1.73(3) \cdot 10^{3}$ \hspace{0.5mm}  &  $1.71(6) \cdot 10^{3}$ \cite{ChenSavage}      & \hspace{-2.5mm} $1.80(13)  \cdot 10^{3}$ \cite{BarnesExp} \hspace{0.5mm} \\
  \hline
    $8.8$ & \hspace{-0.5mm}  $32.8(1)$  \hspace{0.5mm}   & $11$  &  \hspace{0.5mm} $112(1)$  \hspace{0.5mm}   & $4$  & \hspace{0.5mm} $603(9)$  \hspace{0.5mm}   & $2$  & \hspace{0.5mm} $849(16)$ \hspace{0.5mm}   & $3$ & \makecell{ $1.597(27) \cdot 10^{3}$  }    & \makecell{   $1.62 \cdot 10^{3}$~\cite{BirenbaumExp}\\ \hspace{-1mm}   $1.59(6) \cdot 10^{3}$ \cite{ChenSavage}  }        & \hspace{-3.5mm} $1.586(11)  \cdot 10^{3}$ \cite{BirenbaumExp} \hspace{0.5mm} \\
  \hline
   $9$  & \hspace{-0.5mm}  $31.7(1)$ \hspace{0.5mm}  & $11$  &  \hspace{0.5mm} $105(2)$  \hspace{0.5mm}   & $3$  & \hspace{0.5mm} $591(9)$  \hspace{0.5mm}   & $2$  & \hspace{0.5mm} $829(16)$ \hspace{0.5mm}     & $3$ & \makecell{ $1.557(27) \cdot 10^{3}$  }     & \makecell{  $1.58 \cdot 10^{3}$~\cite{BirenbaumExp} \\ \hspace{-1mm}   $1.55(6) \cdot 10^{3}$ \cite{ChenSavage}  }     & \hspace{-3.5mm} $1.570(36)  \cdot 10^{3}$ \cite{BirenbaumExp} \hspace{0.5mm} \\
   \hline
    $10$  & \hspace{-0.5mm}  $27.1(3)$ \hspace{0.5mm}   & $11$   & \hspace{0.5mm} $89(2)$ \hspace{0.5mm}    & $3$   & \hspace{0.5mm} $535(10)$ \hspace{0.5mm}    & $2$ & \hspace{0.5mm} $730(14)$ \hspace{0.5mm}    & $6$ & \makecell{ $1.382(25) \cdot 10^{3}$    }    & \hspace{-3.5mm} $1.387 \cdot 10^{3}$ \cite{Partovi1964}  & \hspace{-2mm} $1.409(42)  \cdot 10^{3}$ \cite{GraeveExp} \hspace{0.5mm} \\
       \hline
   $11$  & \hspace{-0.5mm}  $23.4(3)$ \hspace{0.5mm}    & $11$   & \hspace{0.5mm} $77(1)$ \hspace{0.5mm}     & $3$ & \hspace{0.5mm} $488(10)$ \hspace{0.5mm}    & $2$   & \hspace{0.5mm} $657(11)$ \hspace{0.5mm}      & $3$  & \makecell{ $1.245(22) \cdot 10^{3}$  }    & -  &   \hspace{-2mm} $1.278(38) \cdot 10^{3}$ \cite{GraeveExp} \hspace{0.5mm} \\
   \hline
      $11.39$  & \hspace{-0.5mm}  $22.2(3)$ \hspace{0.5mm}    & $11$    &  \hspace{0.5mm}  $72(1)$ \hspace{0.5mm}      & $3$    &  \hspace{0.5mm}  $478(8)$ \hspace{0.5mm}     & $3$ & \hspace{0.5mm} $629(10)$ \hspace{0.5mm}      & $3$ & \makecell{ $1.201(19) \cdot 10^{3}$  }       &   $1.19 \cdot 10^{3}$~\cite{BirenbaumExp}    &  \hspace{-1.8mm} $1.257(36)  \cdot 10^{3}$ \cite{BirenbaumExp} \hspace{0.5mm}     \\
   \hline
     $12$  &\hspace{-0.5mm}  $20.5(3)$ \hspace{0.5mm}    & $11$   & \hspace{0.5mm} $66(1)$ \hspace{0.5mm}     & $3$   & \hspace{0.5mm} $454(7)$ \hspace{0.5mm}    & $3$ & \hspace{0.5mm} $589(8)$ \hspace{0.5mm}  & $3$ & \makecell{ $1.129(17) \cdot 10^{3}$  }       & -  &  \hspace{-2mm} $1.161(35) \cdot 10^{3}$ \cite{GraeveExp} \hspace{0.5mm}\\
  \hline
     $12.5$  & \hspace{-0.5mm}  $19.2(2)$ \hspace{0.5mm}  & $11$    &   \hspace{0.5mm} $61.2(6)$ \hspace{0.5mm}     & $3$    &  \hspace{0.5mm} $437(7)$ \hspace{0.5mm}    & $3$  & \hspace{0.5mm} $558(7)$ \hspace{0.5mm} & $3$ &  \makecell{ $1.08(2) \cdot 10^{3}$ }    & -  &  \hspace{-2.5mm} $1.04(10)  \cdot 10^{3}$ \cite{BarnesExp} \hspace{0.5mm} \\
  \hline
     $13$  & \hspace{-0.5mm}  $18.0(2)$ \hspace{0.5mm}    & $11$   &  \hspace{0.5mm} $57.0(5)$ \hspace{0.5mm}  & $3$ & \hspace{0.5mm} $421(6)$ \hspace{0.5mm}   & $3$   &    \hspace{0.5mm} $530(5)$ \hspace{0.5mm}  & $3$  & \makecell{ $1.025(12) \cdot 10^{3}$    }       & -   &  \hspace{-1.5mm} $1.058(32) \cdot 10^{3}$ \cite{GraeveExp} \hspace{0.5mm} \\
   \hline
       $14$  & \hspace{-0.5mm}  $15.9(1)$ \hspace{0.5mm}    & $11$   & \hspace{0.5mm} $49.5(3)$ \hspace{0.5mm}  & $3$   & \hspace{0.5mm} $382(7)$ \hspace{0.5mm}     & $2$ &  \hspace{0.5mm} $479(3)$ \hspace{0.5mm}     & $3$ &  \makecell{ $926(11)$   }     & - &  \hspace{2mm} $963(29)$ \cite{GraeveExp} \hspace{0.5mm} \\
         \hline
              $14.7$  & \hspace{-0.5mm}  $14.7(1)$ \hspace{0.5mm}     & $11$   & \hspace{0.5mm} $45.0(2)$ \hspace{0.5mm}  & $3$  & \hspace{0.5mm} $363(7)$ \hspace{0.5mm}     & $2$  & \hspace{0.5mm} $447(2)$ \hspace{0.5mm}      & $3$  & \makecell{ $870(9)$   }      & 900.3~\cite{BernabeiExp}  &  \hspace{2mm} $925(20)$ \cite{BernabeiExp} \hspace{0.5mm} \\
         \hline
            $15$  & \hspace{-0.5mm}  $14.2(1)$ \hspace{0.5mm}    & $11$   &  \hspace{0.5mm} $43.2(1)$ \hspace{0.5mm}   & $3$ & \hspace{0.5mm} $356(6)$ \hspace{0.5mm}   & $2$   & \hspace{0.5mm} $434(3)$ \hspace{0.5mm}     & $3$  &  \makecell{ $847(9)$  }   & -    &  \makecell{\hspace{-1mm} $867(46)$ \cite{AhrensExp} \\ \hspace{0.8mm} $884(27)$ \cite{GraeveExp}   } \\
   \hline
            $16$  & \hspace{-0.5mm}  $12.7(2)$ \hspace{0.5mm}    & $11$   &  \hspace{0.5mm} $37.8(3)$ \hspace{0.5mm}      & $3$   &  \hspace{0.5mm} $333(5)$ \hspace{0.5mm}      & $2$ &  \hspace{0.5mm} $395(4)$ \hspace{0.5mm}    & $3$ &  \makecell{ $779(10)$   }         &  -  & \hspace{0.5mm} $822(25)$ \cite{GraeveExp}\\
              \hline
                 $17$  & \hspace{-0.5mm}   $11.69(2)$ \hspace{0.5mm}   & $13$   &  \hspace{0.5mm} $33.2(4)$ \hspace{0.5mm}    & $3$ & \hspace{0.5mm} $321(6)$ \hspace{0.5mm}   & $4$   & \hspace{0.5mm} $361(5)$ \hspace{0.5mm}     & $3$  &   \makecell{ $727(11)$   }      &  -   & \makecell{\hspace{-1mm} $730(42)$ \cite{SkopikExp} \\ \hspace{0.8mm} $748(22)$ \cite{GraeveExp}   }  \\
   \hline
        $17.6$  & \hspace{-0.5mm}  $11.08(2)$ \hspace{0.5mm}  & $13$    & \hspace{0.5mm}  $30.8(4)$ \hspace{0.5mm}    & $3$    &  \hspace{0.5mm}  $310(6)$ \hspace{0.5mm}     & $4$  &  \hspace{0.5mm}  $342(5)$ \hspace{0.5mm}  & $3$ & \makecell{ $694(12)$   }     & -  &  \hspace{0.5mm} $770(90) $ \cite{BarnesExp} \hspace{0.5mm} \\
  \hline
                        $18$  & \hspace{-0.5mm}    $10.69(3)$ \hspace{0.5mm}  & $13$   &  \hspace{0.5mm} $29.3(4)$ \hspace{0.5mm}    &  $3$ &  \hspace{0.5mm} $297(4)$ \hspace{0.5mm}    & $5$ &  \hspace{0.5mm} $331(6)$ \hspace{0.5mm}    & $3$ &  \makecell{ $668(10)$    }      & -   & \makecell{\hspace{-1mm} $640(31)$ \cite{SkopikExp} \\ \hspace{0.6mm} $716(22)$ \cite{GraeveExp} }\\
                          \hline
                             $19$  & \hspace{-0.5mm}  $9.82(3)$ \hspace{0.5mm}   & $13$   &  \hspace{0.5mm} $25.9(5)$ \hspace{0.5mm}    & $3$ & \hspace{0.5mm} $279(4)$ \hspace{0.5mm}   & $5$   & \hspace{0.5mm} $315(6)$ \hspace{0.5mm}     & $5$  &  \makecell{ $630(10)$   }     & -     &  \makecell{\hspace{-1mm} $637(31)$ \cite{SkopikExp} \\  \hspace{0.8mm} $687(21)$ \cite{GraeveExp}  }\\
   \hline
           $19.3$  &  \hspace{-0.5mm}  $9.58(3)$ \hspace{0.5mm}     & $13$   &  \hspace{0.5mm} $24.5(5)$ \hspace{0.5mm}   & $2$  & \hspace{0.5mm} $274(4)$ \hspace{0.5mm}     & $5$  & \hspace{0.5mm} $308(6)$ \hspace{0.5mm}     & $5$ & \makecell{ $616(11)$    }       & 627.3~\cite{BernabeiExp} &  \hspace{1mm} $617(9)$ \cite{BernabeiExp} \hspace{0.5mm} \\
         \hline
                                        $20$  & \hspace{0.5mm}   $9.05(3)$ \hspace{0.5mm}  & $13$   &  \hspace{0.5mm} $22.9(2)$ \hspace{0.5mm}    & $5$   &  \hspace{0.5mm} $263(5)$ \hspace{0.5mm}   & $5$ &  \hspace{0.5mm} $283(3)$ \hspace{0.5mm}    & $10$ &  \makecell{ $578(8)$    }  &  $588$ \cite{Partovi1964}   & \makecell{\hspace{-1mm} $604(29)$ \cite{SkopikExp} \\ \hspace{-0.8mm} $585(32)$ \cite{AhrensExp}   }\\
  \hline
 \end{tabular}
\label{tab6}
\end{table*}

\end{document}